\documentclass[sigconf, nonacm]{acmart}
\usepackage{subcaption} 
\usepackage{cleveref}
\usepackage{algorithm}
\usepackage{algpseudocode}
\usepackage{array}
\usepackage{tikz}
\newcommand{\circled}[1]{\raisebox{0.5pt}{\textcircled{\raisebox{-0.9pt}{#1}}}}
\usepackage{xcolor}

\newcommand\vldbdoi{XX.XX/XXX.XX}
\newcommand\vldbpages{XXX-XXX}
\newcommand\vldbvolume{14}
\newcommand\vldbissue{1}
\newcommand\vldbyear{2020}
\newcommand\vldbauthors{\authors}
\newcommand\vldbtitle{\shorttitle} 
\newcommand\vldbavailabilityurl{URL_TO_YOUR_ARTIFACTS}
\newcommand\vldbpagestyle{plain}

\usepackage{xcolor}

\newcommand{\ourmethod}{\textsc{SAGE}}

\usepackage{graphicx}
\usepackage{caption}
\usepackage{subcaption}
\usepackage{makecell}
\usepackage[inline]{enumitem}
\usepackage{float}

\begin{document}
\title{
SAGE: Selective Attention-Guided Extraction for Token-Efficient Document Indexing
}

\settopmatter{authorsperrow=4}
\author{Xinzhi Wang}
\affiliation{%
  \institution{Purdue University}
}
\email{wang6171@purdue.edu}

\author{Peter Baile Chen}
\affiliation{%
  \institution{MIT}
}
\email{peterbc@mit.edu}

\author{Gerardo Vitagliano}
\affiliation{%
  \institution{MIT}
}
\email{gerarvit@csail.mit.edu}

\author{Matthew Russo}
\affiliation{%
  \institution{MIT}
}
\email{mdrusso@csail.mit.edu}

\author{Jun Chen}
\affiliation{%
  \institution{Independent Researcher}
}
\email{chjuncn@gmail.com}

\author{Michael Cafarella}
\affiliation{%
  \institution{MIT}
}
\email{michjc@csail.mit.edu}

\author{Samuel Madden}
\affiliation{%
  \institution{MIT}
}
\email{madden@csail.mit.edu}

\author{Chunwei Liu}
\affiliation{%
  \institution{Purdue University}
}
\email{chunwei@purdue.edu}

\begin{abstract}
Large language models with long context windows can answer complex questions directly from full-length academic, technical, and policy documents, but passing entire documents is often costly, slow, and can degrade answer quality while increasing the risk of unnecessary data leakage. This paper targets the common setting of answering many heterogeneous questions over long document(s), where fixed position heuristics and standard retrieval-augmented generation (RAG) can fail due to document structure variability and weak query-chunk semantic similarity, which often requires task- and domain-specific tuning of embedding retrievers. We propose {Selective Attention-Guided Extraction} (\ourmethod), a training-free, plug-and-play context reduction framework that uses a lightweight local LLM to perform a single prefilling pass and convert language model attention signals into a query-specific relevance heatmap at configurable granularities. \ourmethod\ further introduces \emph{differential attention} strategies to better isolate question-relevant evidence, then selects the top-scoring units under a user-defined token budget and forwards only this reduced context to a downstream LLM for answer generation. 
\ourmethod\ surpasses traditional reduction techniques across multiple long-document QA benchmarks, notably securing a top-4 rank on QuALITY-hard while constrained to a 10\% context budget. This enables a 90\% reduction in tokens with competitive accuracy, without the need for model fine-tuning or complex calibration.

\end{abstract}

\maketitle

\pagestyle{\vldbpagestyle}
\begingroup\small\noindent\raggedright\textbf{PVLDB Reference Format:}\\
\vldbauthors. \vldbtitle. PVLDB, \vldbvolume(\vldbissue): \vldbpages, \vldbyear.\\
\href{https://doi.org/\vldbdoi}{doi:\vldbdoi}
\endgroup
\begingroup
\renewcommand\thefootnote{}\footnote{\noindent
This work is licensed under the Creative Commons BY-NC-ND 4.0 International License. Visit \url{https://creativecommons.org/licenses/by-nc-nd/4.0/} to view a copy of this license. For any use beyond those covered by this license, obtain permission by emailing \href{mailto:info@vldb.org}{info@vldb.org}. Copyright is held by the owner/author(s). Publication rights licensed to the VLDB Endowment. \\
\raggedright Proceedings of the VLDB Endowment, Vol. \vldbvolume, No. \vldbissue\ %
ISSN 2150-8097. \\
\href{https://doi.org/\vldbdoi}{doi:\vldbdoi} \\
}\addtocounter{footnote}{-1}\endgroup

\ifdefempty{\vldbavailabilityurl}{}{
\vspace{.3cm}
\begingroup\small\noindent\raggedright\textbf{PVLDB Artifact Availability:}\\
The source code, data, and/or other artifacts have been made available at \url{https://github.com/Tranway1/AttentiveTrim.git}.
\endgroup
}

\section{Introduction}
\label{sec:intro}

As LLM-based applications take on increasingly complex workloads, the demand for long-context question answering has grown rapidly \citep{liu277271533comprehensive, liu2025palimpzest, lin2024towards, ma2024mmlongbench, qian2024streaming, weng2024longvlm, zhang2024long}. Although recent models can ingest entire documents end-to-end \citep{liu277271533comprehensive}, \emph{more context is not always better}. Long inputs often mix a small amount of salient evidence with substantial irrelevant text, which can distract the model and degrade answer accuracy \citep{jin2024long, liu2024lost}. The practical overhead is also substantial: under pay-per-token commercial APIs, processing full documents directly increases monetary cost, and even in self-hosted settings, it wastes compute on non-essential text.

These issues are exacerbated when using smaller local models. Once an input exceeds the context window of a model, the document can no longer be processed in a single forward pass, requiring chunking or multi-stage pipelines that increase latency and memory overhead while complicating deployment.

Beyond efficiency and quality, minimizing unnecessary data exposure is a practical constraint in many real-world settings. Enterprise reports, proprietary documents, and internal analyses often contain sensitive information that should not be fully revealed to downstream or business-facing LLMs. Together, these considerations motivate \emph{context reduction}: extracting a compact subset of the input that preserves answer quality while bounding LLM cost and limiting unnecessary information exposure.

A natural first idea for context reduction is to take advantage of where answers usually appear in a document. Prior work \citep{lin2024towards} and our early observations suggest that some question types do show rough structural patterns. For example, in research papers, answers are often found in sections such as the abstract, introduction, evaluation, or conclusion. This motivates validation-based positional heuristics: using a small set of documents with known answers, we can locate the answer spans, map them to token positions, and build a relevance heatmap over the document. This heatmap can then be used to guide context extraction for new documents.


\begin{figure}[t]
    \centering

    \begin{subfigure}[t]{\linewidth}
        \centering
        \includegraphics[trim=0 0 0 23.5pt, clip, width=0.8\linewidth]{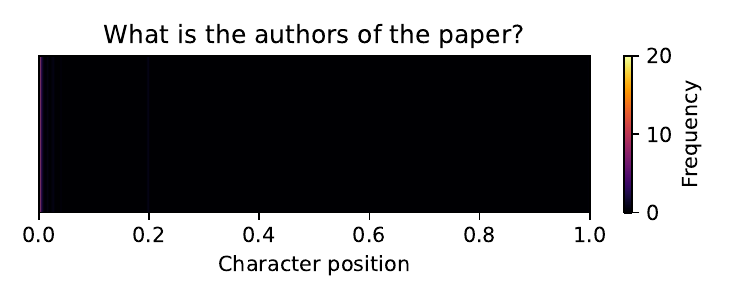}
        \vspace{-1em}
        
        \caption{Who are the authors of the paper}
        \label{fig:ph-a}
    \end{subfigure}

    \begin{subfigure}[t]{\linewidth}
        \centering
        \includegraphics[trim=0 0 0 23.5pt, clip, width=0.8\linewidth]{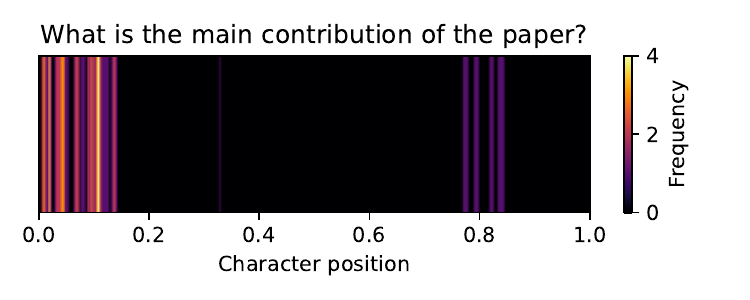}
        \vspace{-1em}
        \caption{What is the main contribution of the paper}
        \label{fig:ph-d}
    \end{subfigure}
        \vspace{-1em}
    \caption{Heatmaps of ground-truth answer distributions in the Paper dataset. Author queries (a) cluster at the start, while contribution queries (b) disperse throughout.}
    \label{fig:position-heatmap}
    \vspace{-2em}
\end{figure}

However, these patterns are only rough, and validation-based heatmaps have clear limitations. First, they rely on a strong LLM to accurately locate answer spans in long documents, which can be unreliable when documents are lengthy or vary in structure. Second, answer locations can differ a lot across question types. As shown in \Cref{fig:position-heatmap}, authorship-related evidence is usually concentrated near the beginning of a paper, while evidence for contributions is spread much more broadly. Third, for reasoning-heavy questions, the relevant evidence is often distributed across several sections rather than confined to a single region. For instance, a paper’s main contribution may be introduced in the abstract, motivated in the introduction, explained in detail later, and summarized again in the conclusion. In these cases, a single average position or fixed region is unlikely to capture all the evidence needed to answer the question correctly.

Retrieval-Augmented Generation offers an alternative by retrieving relevant evidence before generation, reducing the need to pass entire documents to the model. RAG systems span dense bi-encoders as well as token-level late interaction retrievers such as ColBERT \citep{khattab2020colbert} and ColBERTv2 \citep{santhanam2022colbertv2}, with recent refinements that further improve token-level scoring \citep{kang2025trial}. With the emergence of long-context LLMs, RAG has also moved toward retrieving larger and more coherent units \citep{jiang2024longrag} and organizing evidence hierarchically \citep{sarthi2024raptor}. Yet retrieval still relies on embedding similarity, which does not guarantee that selected chunks contain the answer. This gap is most visible when queries lack discriminative keywords or when the supportive context is only weakly aligned with the query in the embedding space. As a result, retrieval may miss critical evidence or include misleading context. For example, for a query such as ``Who are the authors of this paper?'', the chunks most similar in embedding space are often in the reference section (where author names appear in citations) rather than near the title or header, where the correct answer is typically found.
This failure mode becomes especially painful when many questions must be answered over the same document.

These limitations point to a more adaptive, model-driven approach. Instead of inferring relevance indirectly via validation heatmaps or embedding similarity, we ask "\emph{can relevance be obtained directly from the model itself?}"
Transformer-based LLMs compute attention distributions during the prefilling phase that reflect how strongly different parts of the context contribute to answering a given question. Importantly, these signals are available without generating any output. We only run the prefill pass and read the attention weights, avoiding the dominant cost of the decoding phase \citep{agrawal2025evaluating}. We find that these attention signals provide a fine-grained, query-specific relevance signal that can be exploited without external supervision.

Building on this insight, we propose \textbf{S}elective \textbf{A}ttention-\textbf{G}uided \textbf{E}xtraction (\ourmethod), an attention-driven context reduction framework that uses a lightweight local LLM as a preprocessing filter. Given a document and a query, the local model computes token-level attention over the document, producing a relevance heatmap that can be aggregated at configurable granularities such as tokens, sentences, or sections.
We then select the highest-scoring units to form a compact context that can be passed to any downstream LLM for answer generation under a user-provided token budget.

A practical challenge is that raw attention can be noisy: generic document elements such as headers, boilerplate text, and formatting cues may consistently attract attention even when they are not query-relevant. Recent work suggests that contrastive signals can help separate shared background patterns from task-specific evidence \citep{ye2024differential, lim2025grouped, ge2025focusing}. Inspired by this line of work, \ourmethod\ introduces \emph{differential attention}. We subtract attention patterns induced by a contrasting query to cancel persistent noise and better isolate query-specific evidence.

This design offers several system-level advantages. First, \ourmethod\ reduces the original document to a compact context that retains only the content most relevant to the query. Second, it provides explicit budget control via a user-specified token budget, bounding inference cost regardless of document length. Third, unlike validation-based heatmap construction, \ourmethod\ does not require ground truth answers, validation datasets, or query-specific calibration. More broadly, the approach is training-free and can be applied directly with off-the-shelf LLMs, without model modification or fine-tuning, making it readily applicable to new questions and domains. In addition, \ourmethod\ leverages KV-cache reuse to improve efficiency by avoiding repeated encoding of the same document chunks across queries.

We evaluate \ourmethod\ on four benchmarks spanning different modalities. On QuALITY-hard \citep{pang2022quality}, \ourmethod\ ranks \textbf{4th} on the public leaderboard \citep{QuALITYLeaderboard} while using only a \textbf{10\%} context budget. On \textit{Paper} and \textit{Notice} \citep{lin2024towards}, \ourmethod\ consistently outperforms strong embedding-based RAG baselines across token budgets, and we analyze the effects of attention model size and differential-attention variants. Finally, we show that \ourmethod\ extends to semi-structured tables without dataset-specific tuning, achieving accuracy comparable to full-table input while substantially reducing the table content passed to the generator. Looking ahead, we expect the same attention-guided principle to extend to additional modalities, including more complex semi-structured tables and visual documents.




To summarize, our contribution includes:
\begin{itemize}[leftmargin=*, nosep]
    \item We study \emph{budgeted context reduction} in the setting of \emph{multiple questions per document}, and highlight limitations of validation-based positional heuristics and embedding only retrieval for dispersed evidence and weak lexical cues (Section~\ref{sec:intro}).

    \item We propose \textbf{S}elective \textbf{A}ttention-\textbf{G}uided \textbf{E}xtraction (\ourmethod), a training-free pipeline that uses a lightweight local LLM to compute token-level relevance and select compact context under an explicit token budget (Section~\ref{sec:overview}).

    \item We introduce \emph{differential attention} to filter out query-independent noise and improve relevance estimation, and study alternative reference queries, including a fixed baseline prompt and a farthest question contrast (Section~\ref{sec:differential-attention}).

    \item We evaluate \ourmethod\ on four benchmarks spanning different regimes and modalities, showing consistent improvements over strong baselines in token accuracy trade-offs (Section~\ref{sec:evaluation}).

    \item We demonstrate an extension of \ourmethod\ to table selection, achieving accuracy comparable to full-table input while reducing the \emph{average row usage} exposed to the generator (Section~\ref{sec:results-aitqa}).
\end{itemize}


\section{Related Work}
\label{sec:related}

To mitigate the high overhead of long-context LLM inference, prior work spans several distinct but complementary directions. 

\subsection{Long-Context RAG}
\label{sec:related-longcontext-rag}

Retrieval-Augmented Generation (RAG) avoids full document ingestion by conditioning generation only on retrieved evidence. While early systems relied on dense single-vector encoders, late interaction retrievers like ColBERT \citep{khattab2020colbert}, ColBERTv2 \citep{santhanam2022colbertv2}, and TRIAL \citep{kang2025trial} improve localization by retaining token-level structure. As long-context LLMs emerged, methods like LongRAG \citep{jiang2024longrag} and RAPTOR \citep{sarthi2024raptor} adapted by retrieving larger, coherent multi-thousand-token units or hierarchical summaries.

However, simply retrieving more chunks is not a panacea. \citet{jin2024long} shows that increasing the number of retrieved passages can actively distract the reader model with ``hard negatives'', chunks of text that are highly similar to the query in embedding space but actually contain irrelevant or misleading information. This risk of degraded reasoning motivates the need for refinement pipelines like LongRefiner \citep{jin2025hierarchical} and OP-RAG \citep{yu2024defense}.

\ourmethod\ is highly complementary to these directions. While RAG pipelines select candidate documents or chunks, \ourmethod\ acts as a budget-constrained filter to preserve only the highest-utility evidence \emph{within} those texts with a finer granularity.

\subsection{Prompt Compression}
\label{sec:related-prompt-compression}

When retrieval is insufficient, \emph{prompt compression} reduces the effective prompt length while preserving necessary information \citep{li2025prompt}. One approach compresses text into continuous \emph{soft prompts} \citep{li2025prompt} using learned vectors or reusable tokens, as seen in CC \citep{wingate2022prompt} and GIST \citep{mu2024learning}. ICAE \citep{ge2023context} distills text using a frozen LLM, xRAG \citep{cheng2024xrag} injects document embeddings via a fusion bridge, and MIPRO \citep{opsahlong2024optimizing} optimizes prompt programs via meta-optimization. However, soft prompts are opaque, require domain-specific training \citep{li2025prompt}, and often cause semantic distortion \citep{liskavets2025prompt}.

To avoid meaning distortion, \emph{hard prompt} methods select or prune discrete tokens. Early methods like PoWER-BERT \citep{goyal2020power} and TR-BERT \citep{ye2021tr} prune tokens based on learned signals but often adopt fixed lengths on training data. Adaptive methods tailor pruning dynamically. SpAtten \citep{wang2021spatten} uses proportional lengths, LTP \citep{kim2022learned} applies learned thresholds, and LLMLingua\citep{jiang2023llmlingua} scores and removes low-utility tokens. LongLLMLingua \citep{jiang2024longllmlingua} further mitigates the risk of dropping crucial evidence by adding question-aware signals. Similarly, task-specific importance metrics are now actively used to compress dense token streams in multimodal settings \citep{shaosurvey}.

To prevent the fragmentation of local syntax caused by token dropping, recent \emph{context reduction} systems extract contiguous sentences. CPC \citep{liskavets2025prompt} retains top-ranked sentences using a context-aware encoder, while ZenDB \citep{lin2024towards} leverages hierarchical structure in templated documents. Unlike aggressive compressors that suffer from knowledge overwriting \citep{guo2026less}, \ourmethod\ strictly preserves the original source text, avoiding structural assumptions by relying solely on attention signals to extract contiguous windows without training.

\subsection{KV Cache Compression}\label{sec:related-kv-compression}A complementary line of work reduces memory and latency during decoding by optimizing the KV cache without altering the input prompt. FastGen \citep{ge2023model} and SnapKV \citep{li2024snapkv} profile attention patterns to compress caching for heavy-hitting tokens. For dynamic cache management, $H_2O$ \citep{zhang2024h2o} formulates KV eviction as a submodular optimization problem, while NACL \citep{chen2024nacl} improves eviction robustness via proxy-token estimation. Beyond eviction, RetrievalAttention \citep{liu2024retrievalattention} manages cache access via approximate nearest-neighbor search, and SepLLM \citep{chen2024sepllm} leverages separator tokens as compact anchors. Other optimizations include dynamic cache budget allocation across layers \citep{zhou2024dynamickv} and concatenating precomputed caches across repeated contexts, as seen in KVLink \citep{yang2025kvlink}. While effective at reducing decoding costs, most KV-cache methods still require ingesting the full context during the computationally expensive prefilling stage. In contrast, \ourmethod\ reduces the input context \emph{before} generation, lowering both prefilling and decoding costs while providing explicit budget control and supporting privacy-preserving deployments.

\subsection{Differential and Contrastive Attention}
\label{sec:related-differential-attn}

Several studies highlight the critical role of attention structure for extractive tasks and long-context reasoning \citep{wu2024retrieval, zhang2024found}. Historically, differential attention was originally introduced in image processing via the Differential Attention Network (DAN), which contrasts supporting and opposing exemplars to better align model focus with human attention \citep{patro2018differential}. In modern LLMs, methods like Differential Transformer \citep{ye2024differential} and Grouped Differential Attention \citep{lim2025grouped} introduce fundamental architectural changes, computing attention as the difference between softmax maps to cancel common-mode noise. Contrastive signals are also leveraged to isolate task-relevant visual regions in multimodal models \citep{ge2025focusing} and to suppress generic text generation during the decoding phase \citep{li2023contrastive}.

In contrast to approaches that require heavy model modifications, specialized training regimens (e.g., DAN \citep{patro2018differential}), or decoding-time interventions, \ourmethod\ uses differential reasoning as a training-free filter for input context reduction. By applying this contrastive signal purely during the prefilling stage without altering the underlying Transformer, \ourmethod\ provides a lightweight, plug-and-play solution to neutralize query-agnostic noise and isolate discriminative evidence prior to generation.

\section{Methodology}
\label{sec:overview}
\begin{figure*}[t]
    \centering
    \vspace{-2.5em}\includegraphics[width=\textwidth]{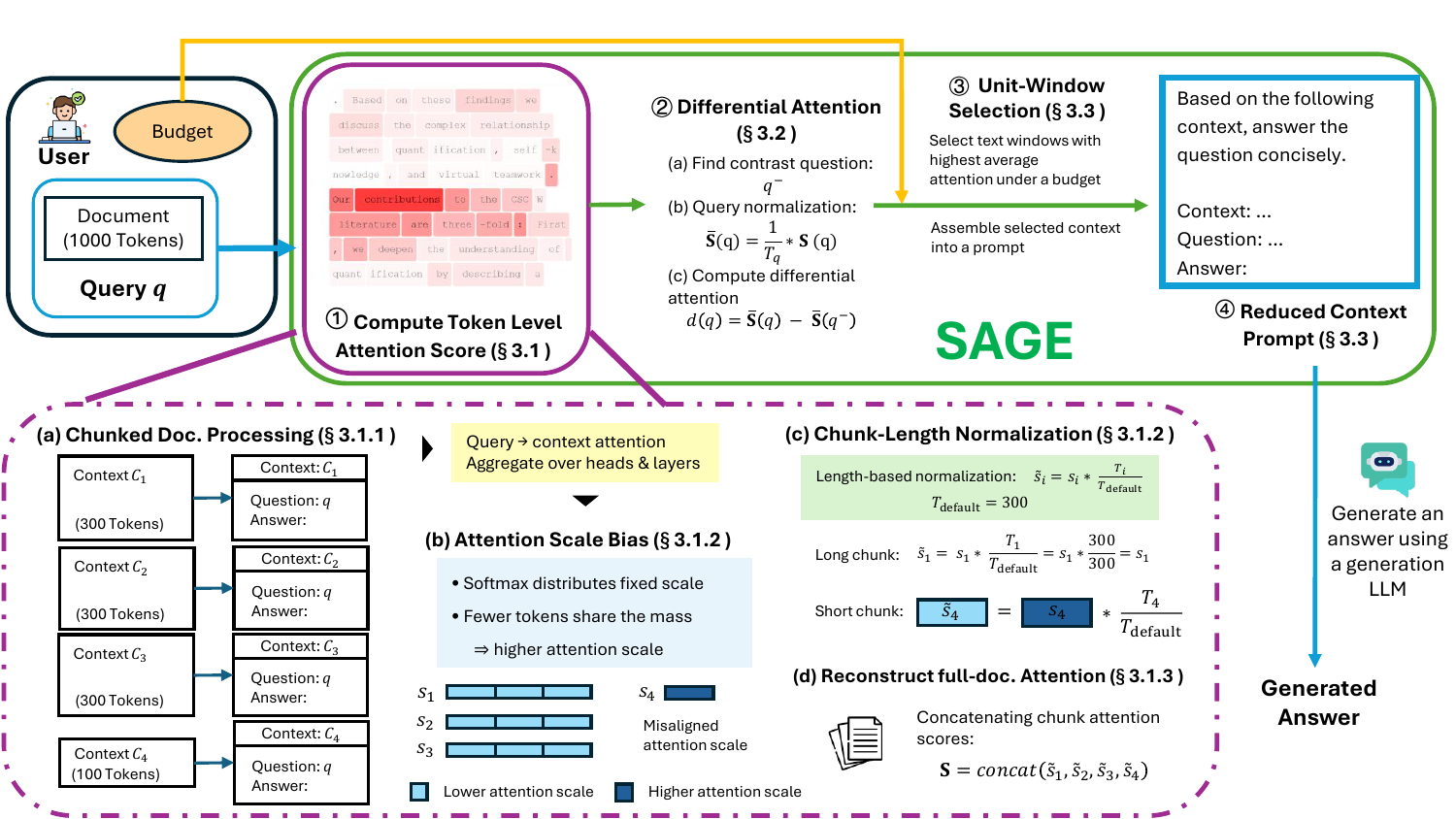}
    \caption{
    \textbf{Overview of the \ourmethod\ pipeline.} \circled{1} \emph{Attention computation} utilizes chunked processing and length normalization to form a document-level score. \circled{2} \emph{Differential attention} refines relevance by filtering structural noise. \circled{3} \emph{Unit-window selection} extracts top text spans under the budget. \circled{4} Spans are assembled into a \emph{reduced-context prompt} for final answer generation.
 }
    \label{fig:overview}
    \label{fig:chunk_normalization}
\end{figure*}


Building upon the need for efficient, noise-resistant context management, this section details the architecture of \ourmethod. Our primary design objective is to extract the most informative text spans from a long document while strictly adhering to a user-specified token budget, ensuring that the generation LLM receives a high-quality, continuous context.

As illustrated in \Cref{fig:overview}, the \ourmethod\ pipeline processes a user-provided \textbf{document}, \textbf{query}, and \textbf{token budget} across four primary stages. Stage \circled{1} computes a query-aware document attention score through four substeps. First, in step (a), we partition the document into chunks that fit within the context window of a lightweight local LLM, computing token-level attention scores for each chunk paired with the query. Because these raw scores can be skewed by chunk length, we identify attention scale misalignment across chunks in step (b) and apply \emph{chunk-length normalization} in step (c) to remove the scale difference caused by varying chunk sizes. Finally, in step (d), we map these normalized scores back to their original positions to reconstruct a cohesive, full document attention score.

In Stage \circled{2}, we first apply \emph{query normalization} so scores are on a consistent scale across different questions. Then, \emph{differential attention} is used to filter out noise and highlight query-relevant signals. {Unit-window selection} (Stage \circled{3}) leverages these refined scores to extract the most useful, coherent text spans within the allocated budget. Finally, in Stage \circled{4}, these selected spans are assembled into a reduced-context prompt and passed to the generation model to produce the final answer. In the following subsections, we discuss each stage in detail.


\subsection{Chunked Attention Computation and Normalization}
\label{sec:attention-computation}

\subsubsection{Chunked Attention Computation}

As we try to extract the parts of a long document that are truly relevant for answering a query, we first need a query-aware signal of token importance. Self-attention offers a natural choice. When a Transformer processes the question, each query token allocates attention to the earlier context tokens it relies on. We therefore use a lightweight local LLM to measure how much attention the query assigns to each document token, yielding an efficient relevance score for guiding later selection. Given a document $D$ and a query $q$, let $T_D$ and $T_q$ denote the number of tokens in the tokenized document and query, respectively.


A direct forward pass over the concatenation of the full document and query is often infeasible, because $T_D + T_q$ may exceed the context window of the model. We therefore partition the document into $n$ contiguous chunks:
\[
D = \{C_1, C_2, \ldots, C_n\}.
\]
Each chunk $C_i$ contains $T_i$ tokens, where $T_i \le T_{\text{default}}$, and $T_{\text{default}}$ is a predefined chunk size chosen to fit the local model context window. In practice, all chunks except possibly the last one have length $T_{\text{default}}$.

For each chunk $C_i$, we construct the prompt
\[
\texttt{Context: } C_i \ ;\ \texttt{Question: } q \ ;\ \texttt{Answer:}
\]
and run a forward pass through the local LLM to extract self-attention scores. Such a forward pass only requires prefilling, and no decoding or token generation is performed. This is sufficient, as our goal is to read off how the model routes attention when interpreting the question, not to produce an answer at the current stage. As the context is placed before the query in the input sequence, each query token can attend to all tokens in $C_i$. We aggregate these attention scores across query tokens, layers, and heads to obtain a relevance score for each context token.


For the $t$-th token in chunk $C_i$, the raw attention score is
\begin{equation}
    s_{i,t}
    \;=\;
    \sum_{k=1}^{T_q}
    \sum_{\ell \in \mathcal{L}}
    \sum_{h \in \mathcal{H}}
    A_{\ell,h}\!\left(q_k, c_{i,t}\right),
    \label{equ:raw_attention}
\end{equation}
where $c_{i,t}$ denotes the $t$-th token in chunk $C_i$, and $A_{\ell,h}(q_k,  c_{i,t})$ is the self-attention scores assigned to $c_{i,t}$ when processing query token $q_k$ at layer $\ell$ and head $h$.

However, we cannot use the raw score $s_{i,t}$ directly. Chunks might have different lengths, most notably the last chunk, as the example shown in \Cref{fig:overview} Step \circled{1} section (a) and (b). This variation changes the scale of the resulting attention values. We therefore need normalization.


\subsubsection{Chunk-Length Normalization}
In Transformer self-attention, the scores for each query token are produced by a softmax over all visible previous tokens. As a result, each query token distributes a fixed total attention mass across the available context tokens.
When we aggregate attention across query tokens, layers, and heads to form $s_{i,t}$, the attention score 
is influenced not only by token relevance but also by how many tokens share this fixed attention mass.

This dynamic introduces a structural bias during chunked processing. If two chunks are equally relevant but differ in length, the shorter chunk will inherently exhibit a higher attention scale simply because the same total attention mass is distributed across fewer tokens. As a result, tokens in shorter chunks can appear artificially inflated in importance compared to those in longer chunks.

To make the attention scale aligned across chunks, we apply a simple chunk-length based normalization:
\begin{equation}
    \tilde{s}_{i,t}
    \;=\;
    s_{i,t}\cdot \frac{T_i}{T_{\text{default}}},
    \label{equ:chunk_normalization}
\end{equation}
where $T_i$ is the length of chunk $C_i$ and $T_{\text{default}}$ is the default chunk length.

This normalization reduces the scores of shorter chunks in proportion to their length, preventing the inflation caused by higher attention scale. As illustrated in \Cref{fig:overview} Stage \circled{1}, Step (a) and (b), the normalization is especially important for the final chunk, which is often shorter than $T_{\text{default}}$.

\subsubsection{Reconstructing Full Document Attention}
After computing normalized attention scores for all chunks, we reconstruct a full document attention score by concatenating chunk-level vectors in their original order. Let $\tilde{\mathbf{s}}_i$ denote the normalized attention vector for chunk $C_i$. We define the full document attention vector as
\begin{equation}
    \mathbf{S} =
    \text{concat}\left(
    \tilde{\mathbf{s}}_1,\,
    \tilde{\mathbf{s}}_2,\,
    \ldots,\,
    \tilde{\mathbf{s}}_n
    \right).
    \label{equ:doc_attention}
\end{equation}

The resulting vector $\mathbf{S}$ is aligned with the original document token order and serves as the full document relevance score for the next stage.

\subsubsection{KV-cache reuse for efficient attention computation.}
Computing attention independently for every chunk–query pair would require repeatedly encoding the same chunk content for different queries. To eliminate this redundancy, we reuse the model’s key–value (KV) cache.

We decompose the chunk–query input into two parts. The \emph{prefix} is the context section, while the \emph{suffix} is the query section. The query text is inserted after the \texttt{Question:} marker and varies across queries.

For each chunk $C_i$, we encode the prefix once and store the resulting KV states. These cached states represent the hidden representations of the chunk tokens and can be reused for multiple queries. We index cache entries using a hash of the chunk token IDs and maintain the cache per document. When switching to a new document, all cached states are cleared to prevent cross-document reuse.

When processing a query $q$ for chunk $C_i$, we retrieve the cached KV states and run the model only on the query suffix, supplying the cached states as past key values. During this step, each query token attends to the cached context tokens, allowing the model to produce the same attention scores as if the full prefix–suffix sequence had been processed jointly. We then aggregate these attention scores to compute $s_{i,t}$ without re-encoding the chunk.

This cache-based execution preserves the exact attention computation while substantially improving efficiency. Each chunk is encoded once and can be reused across many queries, reducing both computation and memory overhead.

\subsection{Differential Attention}
\label{sec:differential-attention}

\begin{figure}[t]
    \centering

    \begin{subfigure}[t]{0.49\linewidth}
        \centering
        \begin{tikzpicture}
            \node[anchor=south west, inner sep=0] (img) {\includegraphics[width=\linewidth]{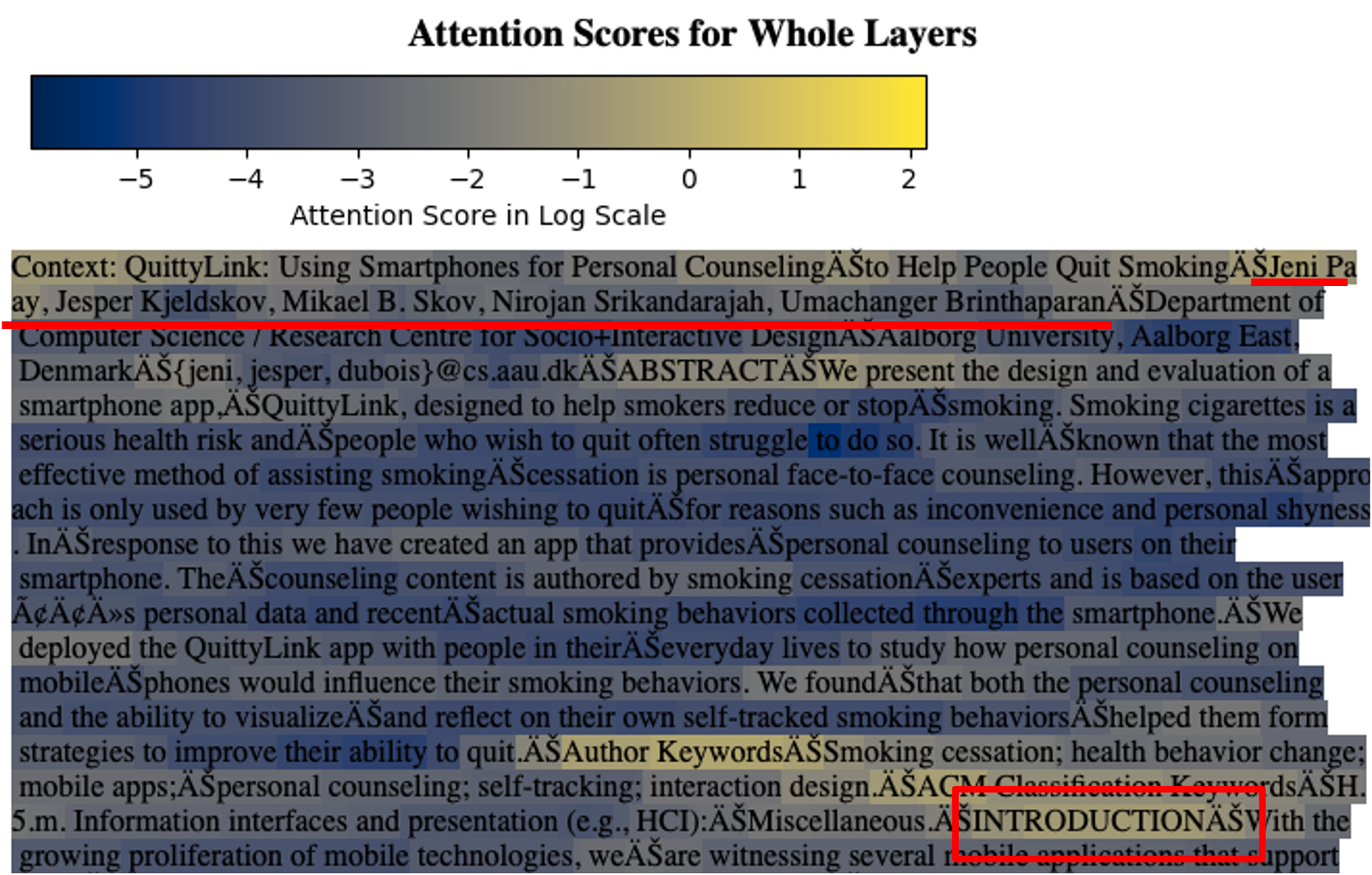}};
        \end{tikzpicture}
        \caption{Raw attention without differential attention}
        \label{fig:ta-a}
    \end{subfigure}
    \hfill
    \begin{subfigure}[t]{0.49\linewidth}
        \centering
        \begin{tikzpicture}
            \node[anchor=south west, inner sep=0] (img) {\includegraphics[width=\linewidth]{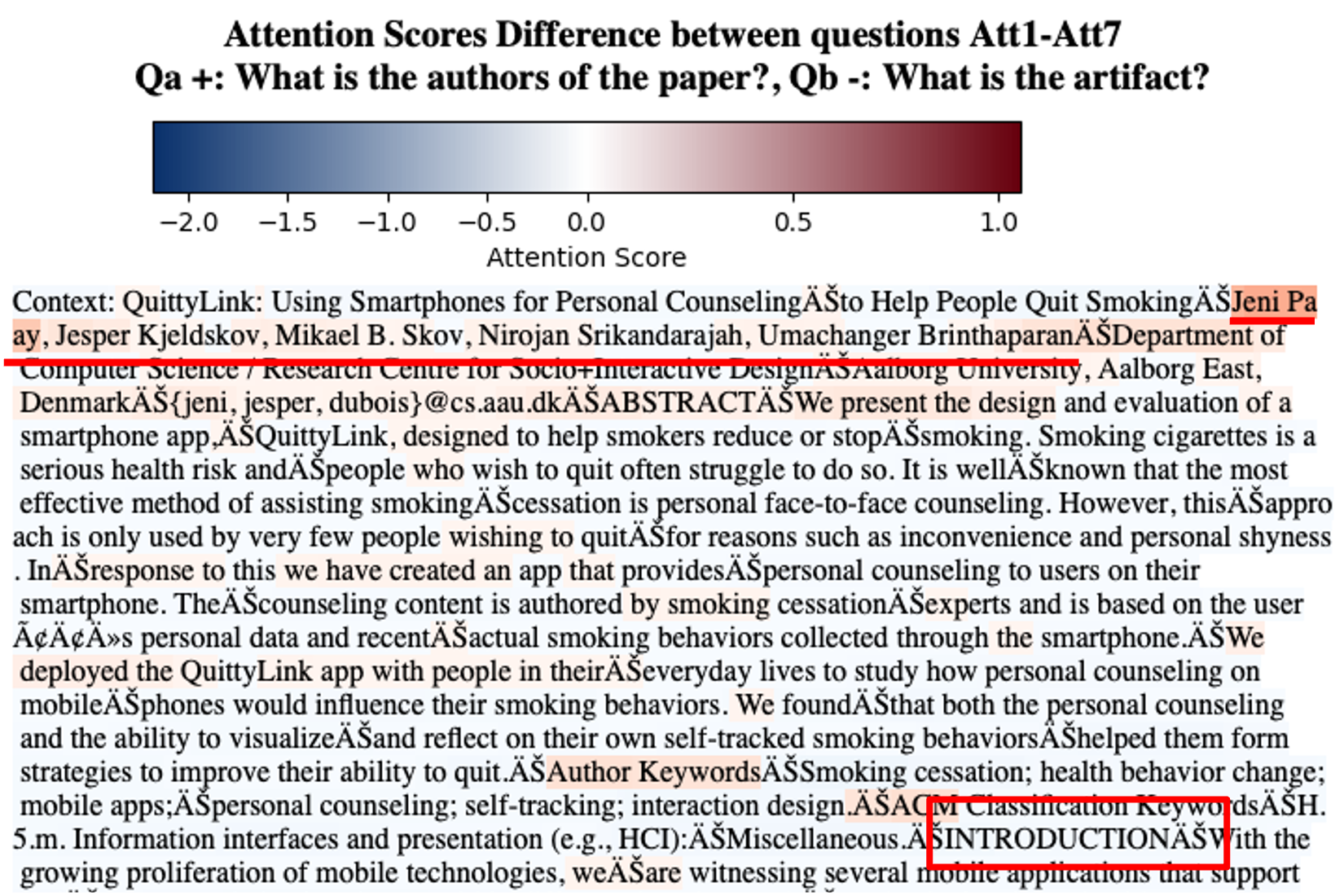}};
        \end{tikzpicture}
        \caption{Differential attention with its farthest query}
        \label{fig:ta-d}
    \end{subfigure}
    \vspace{-1em}
    \caption{
    \textbf{Effect of differential attention on token-level relevance.}
    (a) Without differential attention, generic structural tokens "Introduction" receive high attention. (b) With differential attention, generic content attention is canceled, reducing noise in the attention score.
    }
    \label{fig:token-attention}
\end{figure}


A full document attention map is not yet sufficient for context selection. Raw attention is a useful relevance score, but it often highlights content that is broadly noticeable rather than truly informative for the target query, such as section headers and formatting markers. \Cref{fig:token-attention} illustrates this behavior. In \Cref{fig:ta-a}, the model correctly attends to useful evidence, but it also gives high attention to the header token ``Introduction,'' which is noticeable in the document structure but not necessarily relevant to the query. Because of this, attention-guided selection may waste budget on text that looks important but is not actually useful for answering the question. This observation suggests a simple way to separate signal from background. Many of these high-attention regions are query invariant. They receive attention for a wide range of questions because they are prominent or central to the document structure. If we can estimate this shared background attention, we can subtract it away and keep what remains distinctive to the target query. 

As illustrated in \Cref{fig:ta-d}, we compute attention for the target query and for a contrasting query, then take their difference. Regions that are highly attended under both queries are treated as background and largely cancel out; regions that matter primarily for the query remain prominent, yielding a cleaner, more query-specific relevance score.

This framing raises two practical questions that we address next. First, how to perform the subtraction properly? Second, how do we construct an effective contrasting query that captures document-wide noise while minimally overlapping with the true evidence needed for the target query?


One practical challenge is that we cannot directly subtract the two aggregated attention scores when the queries have different lengths. Each query token spreads a fixed amount of attention mass over the document, so a longer query naturally accumulates a larger total attention magnitude even when the overall attention pattern is similar. To make the two scores comparable, we first normalize attention by query length before taking the difference.

\subsubsection{Query Normalization}
\begin{figure}
    \centering
    \includegraphics[width=\linewidth]{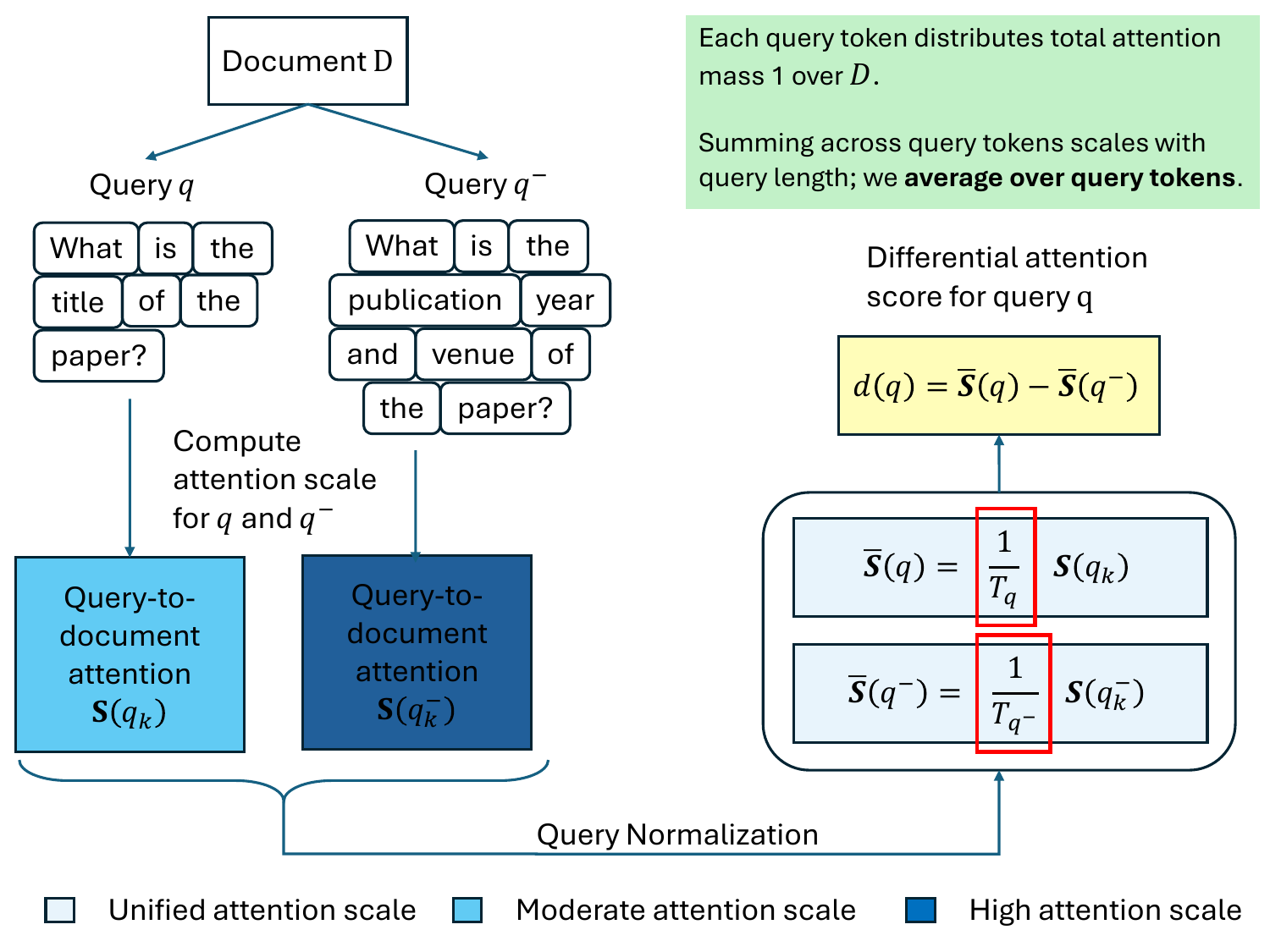}
    \vspace{-2.5em}
    \caption{ \textbf{Query normalization for differential attention.} Each query token distributes a fixed total attention scale over $D$, $\mathbf{S}(q)$ is the aggregation over all heads and layers across tokens with query length. We therefore average the Query-to-document attention scale to obtain $\mathbf{\bar{S}}(q)=\frac{1}{T_q} \mathbf{S}(q)$ and $\mathbf{\bar{S}}(q^-)=\frac{1}{T_{q^-}} \mathbf{S}(q^-)$, then compute $d(q)=\mathbf{\bar{S}}(q)-\mathbf{\bar{S}}(q^{-})$. 
    }
    \label{fig:query_normalization}
        \vspace{-2em}
\end{figure}

\Cref{fig:query_normalization} illustrates why query normalization is necessary. Consider two queries over the same document: a shorter target query $q$, "What is the title of the paper?" and a longer contrasting query $q^{-}$, "What is the publication year and venue of the paper?". Even if both queries attend to the same generic regions of the document, the longer query contributes more total attention mass simply because it has more query tokens, causing a misalignment in attention scale. As a result, directly subtracting the aggregated scores $\mathbf{S}(q)$ and $\mathbf{S}(q^-)$ would confuse query length with relevance.

To remove this bias, we convert each reconstructed query-to-document attention score into an average per-query-token score. Using the full document attention vector from \Cref{equ:doc_attention}, we write $\mathbf{S}(q)$ and $\mathbf{S}(q^-)$ for the reconstructed attention scores obtained with the target query $q$ and the contrasting query $q^{-}$, respectively. We then normalize each score by its query length:
\begin{equation}
    \bar{\mathbf{S}}(q)
    \;=\;
    \frac{1}{T_q}\mathbf{S}(q).
    \label{equ:query_normalization_q}
\end{equation}

Similarly, for a contrasting query $q^{-}$ with length $T_{q^-}$,
\begin{equation}
    \bar{\mathbf{S}}(q^-)
    \;=\;
    \frac{1}{T_{q^-}}\mathbf{S}(q^-).
    \label{equ:query_normalization_qneg}
\end{equation}

We then define the differential attention score for the target query as
\begin{equation}
    \mathbf{d}(q)
    \;=\;
    \bar{\mathbf{S}}(q) - \bar{\mathbf{S}}(q^-).
    \label{equ:diff_attention}
\end{equation}

This normalization unifies the attention scale of the two queries before subtraction, so the difference reflects relevance rather than query length. In practice, it stabilizes differential attention when query lengths vary and prevents longer queries from dominating the difference purely due to scale.

The resulting score $\mathbf{d}(q)$ highlights regions that are more strongly attended by the target query than by the contrasting query. Tokens that attract attention under both queries tend to cancel out and move closer to zero, while tokens that are uniquely relevant to the target query receive larger positive values.

\subsubsection{Applying Differential Attention in \ourmethod}
\label{sec:diff-attn-application}

To make differential attention effective, the contrasting query should attend to document regions that differ from those needed by the target query. This improves separation between target-specific evidence and shared background attention, and therefore increases the denoising effect of subtraction.

In our implementation, we consider three variants for analysis:
\begin{itemize}[leftmargin=*, nosep]
\item \textbf{Raw query ($q$).}
The user’s original question, without differential subtraction.

\item \textbf{Fixed contrast query.}
A fixed prompt shared across the dataset. For this paper, we used \emph{``Please repeat the context.''}.

\item \textbf{Farthest query ($q^{-}$).}
A query selected to be maximally dissimilar to the target query. When multiple questions are available for the same document, we choose the farthest query as the one whose embedding is most distant from the target query embedding. By single differential attention calculation, we got target-specific evidence revealed for both queries. 
\end{itemize}

We compare these variants empirically in \Cref{sec:results-paper} and show that using a semantically distant contrast query yields a cleaner relevance score and improves downstream selection quality.

\subsection{Unit Window Selection with Budget Control}
\label{sec:unit-window}

Given the unified attention scores from the earlier stages, \ourmethod\ converts token-level relevance signals into a set of coherent text snippets that the generation model can consume under a strict token budget.

Although attention is computed at the token level, selecting only isolated high-scoring tokens produces a fragmented context that is poorly suited for generation. LLMs generally benefit from contiguous spans that preserve local semantics. We therefore perform budget-aware selection over fixed-length windows and merge overlapping windows into coherent snippets. To reduce the effect of spiky token-level scores, we also apply a light local smoothing step before window scoring so that nearby tokens in the same evidence span receive more consistent scores.

We define a user-specified budget $b \in (0,1]$ as the fraction of the document to retain. The total number of selected tokens is constrained to $\lceil bT_D \rceil$. We use a base window ratio $w = 0.02$ (2\% of the document length) and define the window length $L$ as
\[
L =
\begin{cases}
\lceil bT_D \rceil, & \text{if } b < w, \\
\lceil wT_D \rceil, & \text{if } b \ge w.
\end{cases}
\]

When $b < w$, selecting a single window preserves the most relevant contiguous region under the budget. When the budget is larger than $w$, fixing the window size allows \ourmethod to select multiple evidence regions distributed across the document.

We slide a length $L$ window over the full document attention matrix and compute an aggregate score for each window. Windows are ranked by score, and selection proceeds greedily under the budget constraint. If a newly selected window overlaps with previously selected spans, the spans are merged. Otherwise, the window is added as a new span. This process continues until the token budget is exhausted.

This strategy avoids two common failure modes. First, selecting a single long span may introduce substantial irrelevant text while missing important evidence in other parts of the document. At the other extreme, selecting many disjoint high-scoring tokens or tiny spans yields a fragmented context that is difficult for the generation model to use. Unit-window selection balances these extremes by concentrating the budget on high utility regions while preserving local coherence.

Finally, the selected snippets are assembled in an original logical order into a reduced context prompt and then passed to a generation model to produce the final answer.


\section{Evaluation}


\label{sec:evaluation}
We evaluate \ourmethod\ on four datasets that stress attention guided context reduction from different angles, including document length, structure, modality, and question format: QuALITY \citep{pang2022quality}, Paper \citep{lin2024towards}, Notice \citep{lin2024towards}, and AIT-QA \citep{katsis2021aitqa}. We report task performance under varying token budgets and compare \ourmethod\ against RAG baselines on each dataset. For evaluation, we use an LLM as a judge to provide a binary correctness decision. We also compute cosine similarity for all questions to measure semantic alignment between \ourmethod’s generated responses and the ground truth. Together, these metrics provide a complementary view of output quality.

Rather than treating each dataset as an isolated benchmark, we use them to examine complementary aspects of long context understanding, including reasoning across distributed evidence, robustness to domain variation, efficiency under tight token budgets, and adaptability to inputs.

We structure our evaluation around the following guiding questions:

\begin{itemize}
\item[(Q1)] Does \ourmethod\ consistently outperform retrieval-based pipelines that use strong embedding models under aligned token budgets?
\item[(Q2)] Can \ourmethod\ handle reasoning-intensive questions when relevant evidence is hard to find by keywords or semantic similarity?
\item[(Q3)] How does token budget size affect accuracy?
\item[(Q4)] How does the size of the attention model influence context selection quality and accuracy?
\item[(Q5)] How does \ourmethod\ compare with retrieval pipelines in runtime cost and computational overhead?
\item[(Q6)] Does \ourmethod\ remain effective across different domains and document structures without dataset-specific tuning?
\item[(Q7)] How do different differential attention strategies influence context selection quality?
\item[(Q8)] Can \ourmethod\ remain effective under budgets smaller than a single retrieval chunk, where chunk-based methods become inflexible?
\item[(Q9)] Can attention-guided selection extend to other modalities while preserving task accuracy?

\end{itemize}

\subsection{Datasets}
\label{sec:datasets}
The four datasets varied in different modalities, context lengths, and QA formats. They also varied in their primary focus and the number of questions, allowing us to evaluate \ourmethod\ comprehensively across diverse settings. \Cref{tab:dataset_summary} summarizes these key characteristics.



\begin{table*}[t]
\centering
\caption{\textbf{Summary of evaluation datasets.} The four datasets provide complementary test settings across different perspectives.}
    \vspace{-1em}
\label{tab:dataset_summary}
\small
\begin{tabular}{p{1.7cm}p{1.8cm}p{1.3cm}p{1.9cm}p{4cm}p{3cm}}
\toprule
\textbf{Dataset} & \textbf{Modality} & \textbf{Context} & \textbf{QA format} & \textbf{Main focus} & \textbf{\# Questions} \\
\midrule
\textbf{QuALITY-hard} & Text (passage) & Long-context & Multiple-choice & Multi-hop and Reasoning intensive reading comprehension & hard subset of QuALITY \\
\midrule
\textbf{Paper} (ZenDB) & Text (scientific papers) & Very long-context & Open ended QA & Mixed factual and semantic extraction; long-document analysis & 9 per document \\
\midrule
\textbf{Notice} (ZenDB) & Text (regulatory notices) & Long context & Open ended QA & Localized factual extraction under aligned token budgets & 6 per document \\
\midrule
\textbf{AIT-QA} & Table (airline SEC filings) & Semi-structured & QA & Table QA over complex headers and domain-specific content & 515 (over 116 tables) \\
\bottomrule
\end{tabular}
\end{table*}

\textbf{QuALITY} \citep{pang2022quality} is a long passage, multiple-choice benchmark for reading comprehension and reasoning. We evaluate on \textbf{QuALITY-hard}, a difficult subset constructed by labeling questions that annotators could not reliably answer under time-constrained search, but could answer in an untimed setting. As a result, QuALITY-hard emphasizes questions that require integrating evidence across the passage rather than relying on localized keyword matches.

\textbf{Paper} and \textbf{Notice} are derived from ZenDB \citep{lin2024towards} and provide multi-question workloads over the same document. \textit{Paper} contains very long scientific publications with both localized metadata queries and open-ended semantic questions, such as the main contribution. This complexity makes it a challenging setting for retrieval and a highly useful testbed for analyzing differential attention. \textit{Notice} consists of regulatory notices written in formal, repetitive legal language. Its questions target specific fields whose answers are typically localized within the document.

Finally, \textbf{AIT-QA} \citep{katsis2021aitqa} evaluates whether attention-guided selection extends beyond free text to semi-structured tables. It contains 515 questions over 116 tables with complex headers and domain-specific terminology.

\subsection{Baselines}
\label{sec:baseline}

We compare \ourmethod\ against a standard RAG pipeline. We instantiate the RAG baselines with top-performing embedding models from the MTEB leaderboard \citep{muennighoff2023mteb,embeddingbenchmark}
, including \emph{Octen-Embedding-4B}, \emph{Qwen3-Embedding-8B}, and \emph{UAE-Large-V1}. These three models cover a strong range of retrieval quality and computational cost, and represent competitive, widely used choices in modern RAG systems.


Retrieval is performed at the document level. For each query, RAG searches only among chunks belonging to the corresponding document, rather than across the entire corpus. For each document, we tokenize the text and split it into overlapping chunks of fixed length. We compute embeddings for all chunks and retrieve the top-$k$ chunks most similar to the query using cosine similarity. Retrieved chunks are reordered according to their original positions in the document, deduplicated in overlapping regions, concatenated, and then fed to the answer-generation model. To isolate the effect of context selection, we use the same answer generation model and decoding configuration for both RAG and \ourmethod. In particular, generation is conditioned on the selected context and the query, with the temperature fixed to 0.0 to eliminate stochastic variation. We vary $k$ to assess how retrieval depth impacts performance and, when feasible, adjust the resulting prompt length to match \ourmethod’s token budget, enabling fair, token-aligned comparisons.

\subsection{Evaluation Metrics}
\label{sec:evalmetric}

We evaluate answer quality using two complementary criteria: \emph{LLM-based judgment} and \emph{semantic similarity}. These metrics are chosen to reflect the heterogeneous answer formats across datasets, ranging from discrete multiple choice and numeric outputs to long, open-ended textual responses.

\paragraph{LLM-Based Judgment.}
Evaluating long-context QA is challenging because many questions admit multiple valid phrasings, and small surface differences do not necessarily indicate incorrect reasoning. We therefore use an LLM as a judge to apply explicit, task-aware grading rules and produce a binary correctness decision. For QuALITY, answers are discrete, so correctness is evaluated by exact match. For the other datasets, the judge directly decides whether an answer is correct, allowing harmless rephrasing while penalizing missing or incorrect core content.

As answer formats and grading criteria differ across datasets, we design a separate judging prompt for each dataset. Each prompt specifies the expected answer format and the criteria for correctness, and is shared by \ourmethod\ and all baselines within that dataset. To ensure the judge behaves as intended, we manually inspected the prompts and audited a substantial sample of judged instances (about 200 per dataset) to verify that the outputs follow the stated rules.

\paragraph{Semantic Similarity Score.}



Binary correctness can hide meaningful differences among partially correct answers, especially for long, open-ended responses where the format may vary even if the core idea aligns with the ground truth. To better characterize answer quality, we additionally compute a semantic similarity score between the generated answer and the reference. Specifically, we embed both texts and report their cosine similarity in [0,1], which is robust to paraphrasing and reflects partial semantic overlap. We analyze similarity at the distribution level using Cumulative Distribution Functions (CDFs) to compare how often each method produces high-similarity answers under different context budgets.

\subsection{Experimental Setup}
\label{sec:setup}

To enable a fair, controlled comparison, we use separate models for attention computation, retrieval, answer generation, and evaluation, while keeping all other settings and pipeline components fixed. 

\paragraph{Attention Models.}
For \ourmethod, attention scores are computed using three language models of increasing scale: \emph{LLaMA-3.2-1B}, \emph{Qwen3-8B}, and \emph{Qwen3-14B}.
These models are used exclusively for attention estimation and do not participate in answer generation.
This setup allows us to study the impact of model capacity on attention quality while keeping the downstream generation fixed.

\paragraph{Answer Generation and Evaluation.}
All methods, including RAG baselines, use \emph{Gemini-2.5-Flash-Lite} for answer generation. By fixing the generation model and decoding configuration, we isolate the effect of context selection. 
{Unless stated otherwise, \ourmethod\ uses the farthest query differential attention variant when constructing the reduced context for generation.} 
LLM-based evaluation is performed using \emph{GPT-4o-mini} across all datasets.

\paragraph{Embeddings.}
Different embedding models are used at different stages to balance efficiency and semantic fidelity.
For farthest question selection in differential attention, we use \emph{all-MiniLM-L6-v2} due to its low computational cost and reliable relative distance estimates.
For semantic similarity evaluation, we use \emph{Qwen3-Embedding-0.6B}, which provides stronger semantic representations while remaining efficient at scale.


\subsection{QuALITY}
\label{sec:results-quality}

QuALITY-hard \citep{pang2022quality} is designed to stress long-context reasoning. The questions are typically unambiguous, but answering them often requires combining evidence from multiple parts of a passage rather than matching a single keyword-rich context. This makes QuALITY-hard a direct test of whether attention-guided context reduction can preserve reasoning quality when relevant evidence is not reliably surfaced by retrieval similarity. In this section, we focus on answering (Q1) to (Q5).

\begin{figure}[htbp]
    \centering
    \includegraphics[width=1\linewidth]{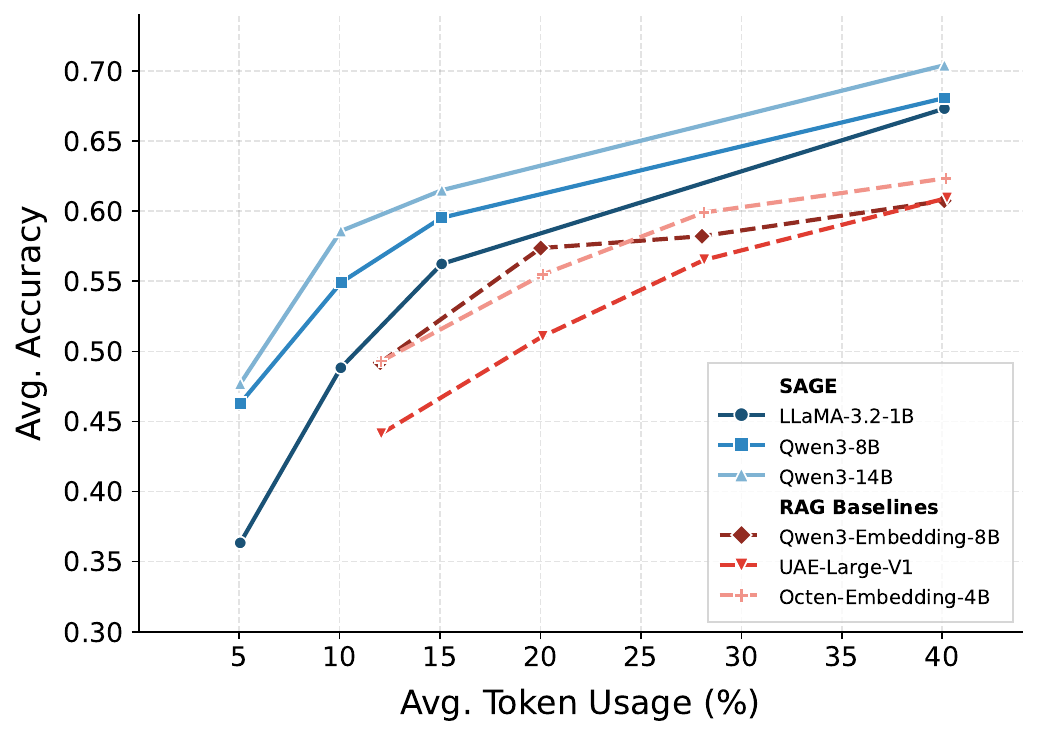}
    \caption{\textbf{Accuracy on QuALITY-hard versus average token usage.} We compare \ourmethod\ using attention models of different sizes with RAG baselines using different retrieval embeddings. Even the 1B attention model outperforms RAG, and performance improves as model size increases.}
    \label{fig:quality}
    \vspace{-2em}
\end{figure}

\subsubsection{Effectiveness compared to strong RAG baselines (Q1, Q2)}
\Cref{fig:quality} shows that \ourmethod\ outperforms on QuALITY-hard across a range of tested token budgets. 
The strength of \ourmethod\ comes from better context selection, not from using a large attention model. For instance, at a 10\% token budget, \ourmethod\ with Qwen3-14B achieves similar accuracy compared to RAG configurations that use nearly four times as many tokens. Moreover, even the smallest attention model, \emph{LLaMA-3.2-1B}, outperforms all strong RAG baselines across the evaluated budget range. \ourmethod\ works well as it selects evidence based on attention signals that reflect how the question connects to different parts of the passage, which helps when evidence is spread across the document.
Overall, these results address (Q1) and (Q2): Across all evaluated settings, \ourmethod\ consistently outperforms RAG baselines at similar or lower token usage, including RAG configurations using strong retrieval embeddings.

\subsubsection{More token budget is not always better (Q3)}
As the token budget increases, both \ourmethod\ and RAG generally improve in \Cref{fig:quality}. However, at larger budgets, the marginal gains diminish, and the curves begin to plateau. Concretely, once the prompt already contains the key evidence, adding more context yields few additional correctly answered questions.  This pattern suggests that QuALITY-hard is not primarily bottlenecked by raw context length. Instead, performance depends largely on retrieval quality and on whether the prompt contains the \emph{right} evidence. The observation is consistent with prior findings that adding more retrieved context does not necessarily translate into better long-context QA accuracy \citep{jin2024long, liu2024lost}. Taken together, these results answer (Q3): accuracy saturates beyond moderate budgets, indicating that evidence selection matters more than simply providing more tokens.

Token usage between \ourmethod\ and RAG is not strictly aligned on QuALITY. QuALITY documents vary substantially in length, approximately 1.5K to 6K words without punctuation \citep{pang2022quality}. Because RAG retrieves fixed-size chunks, the final prompt length depends on the number of retrieved chunks and the amount of overlap among them, which makes it difficult to hit an exact token budget in practice.  In contrast, \ourmethod\ directly optimizes for a user-specified budget and can reliably meet it, while chunk-based retrieval may overshoot or undershoot the desired token limit.

\subsubsection{Ablation on attention model size (Q4)}
As we increase the attention model size from \emph{LLaMA-3.2-1B} to \emph{Qwen3-8B} and \emph{Qwen3-14B}, accuracy improves consistently across budgets. The effect is strongest under tight constraints, where selection mistakes are most costly. At a 5\% token budget, accuracy increases from \textbf{0.36} (1B) to \textbf{0.46} (8B) and \textbf{0.48} (14B), meaning that larger attention models enable \ourmethod\ to answer substantially more questions correctly while operating under the same strict budget. At higher budgets the gap shrinks, again reflecting diminishing returns once most key evidence is already included.

Importantly, \ourmethod\ does not require a large attention model to be effective: even the 1B scorer remains competitive and already outperforms the strongest RAG baselines in \Cref{fig:quality}. Overall, these results answer (Q4), increasing attention model capacity improves selection quality and yields higher QA accuracy, with the largest gains under tight budgets.

\subsubsection{Runtime and computational overhead (Q5)}
We further examine the runtime on QuALITY-hard. For RAG, the dominant overhead comes from computing embeddings for all \textbf{8{,}798} chunks, which takes \textbf{98.2s} (UAE-Large-V1), \textbf{370.7s} (Qwen3-Embedding-8B), and \textbf{838.6s} (Octen-Embedding-4B). For \ourmethod, using \emph{LLaMA-3.2-1B} as the attention model, we answer \textbf{3{,}797} queries in \textbf{690.9s}, with \textbf{3{,}011} cache hits that reuse document encodings across queries. Overall, \ourmethod\ is slower than the UAE and Qwen3 RAG baselines, but faster than the Octen baseline. Taken together, these results answer (Q5). \ourmethod\ operates in a similar runtime regime to retrieval pipelines while achieving materially higher QuALITY-hard accuracy.

\subsubsection{Public leaderboard: competitive ranking under strict budgets}
On the public QuALITY leaderboard \citep{QuALITYLeaderboard}, \ourmethod\ ranks \textbf{4th} on QuALITY-hard while using only a \textbf{10\%} context budget with \emph{Qwen3-8B}. This demonstrates that the accuracy gains we observe in controlled experiments translate to a competitive end-to-end system under strict token constraints. Notably, \ourmethod\ achieves this ranking with substantially lower token usage than many other high-performing systems. Many of those approaches rely on heavier pipelines, such as fine-tuning and longer inputs, which typically lead to higher inference latency.

\subsection{Paper}
\label{sec:results-paper}

\begin{figure*}[htbp]
    \centering
    \begin{subfigure}[t]{\linewidth}
        \centering
        \includegraphics[width=\linewidth]{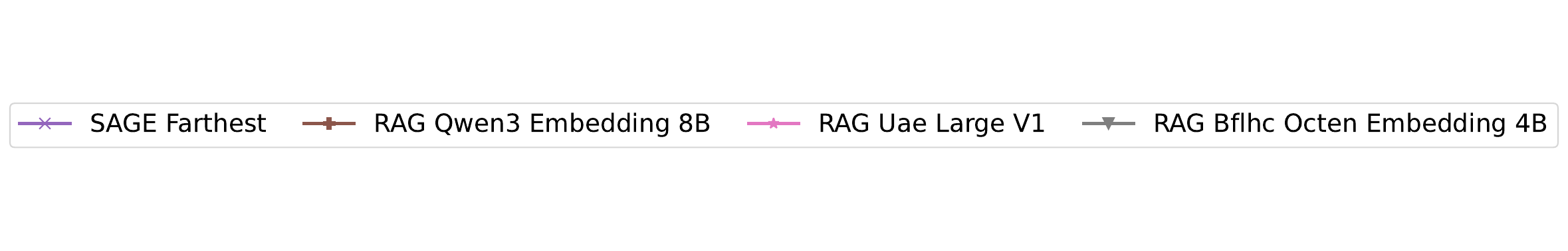}
        \label{fig:lpr-0}
    \end{subfigure}
    \vspace{-40pt}

    \begin{subfigure}[t]{0.24\textwidth}
        \centering
        \includegraphics[width=\linewidth]{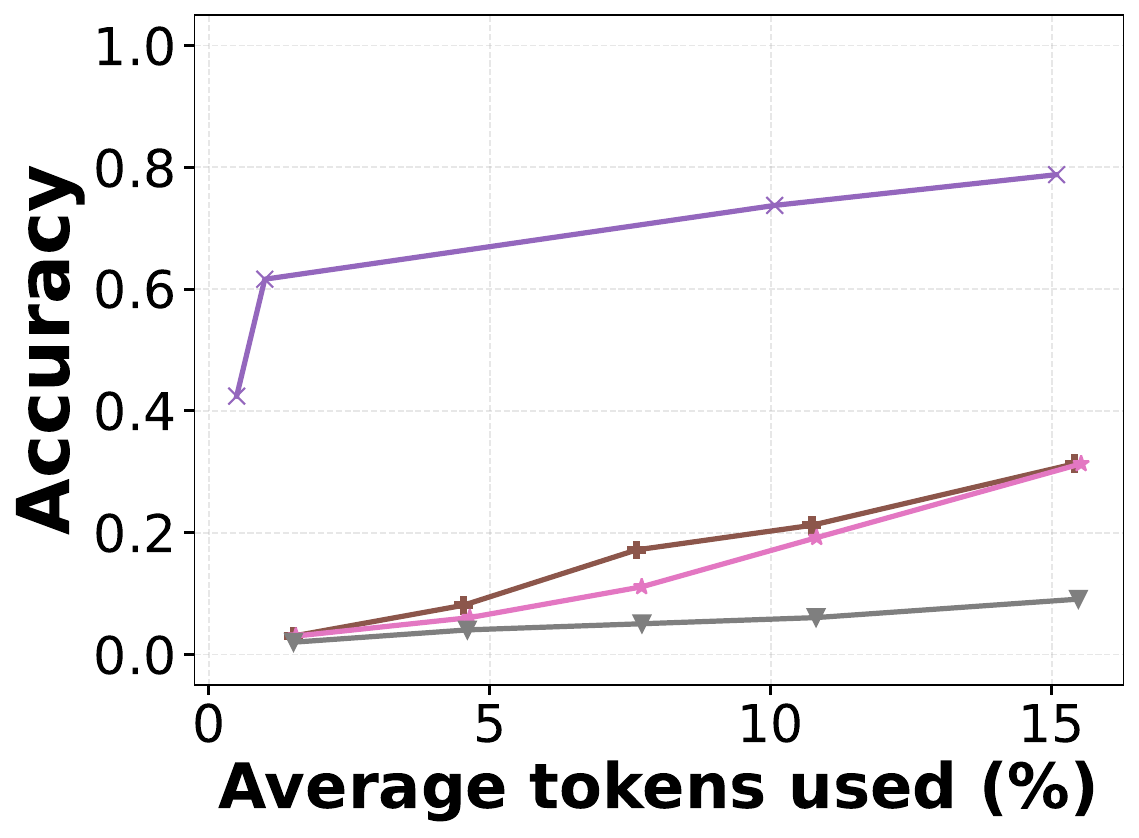}
        \caption{What is the number of authors}
        \label{fig:lpr-a}
    \end{subfigure}
    \begin{subfigure}[t]{0.24\textwidth}
        \centering
        \includegraphics[width=\linewidth]{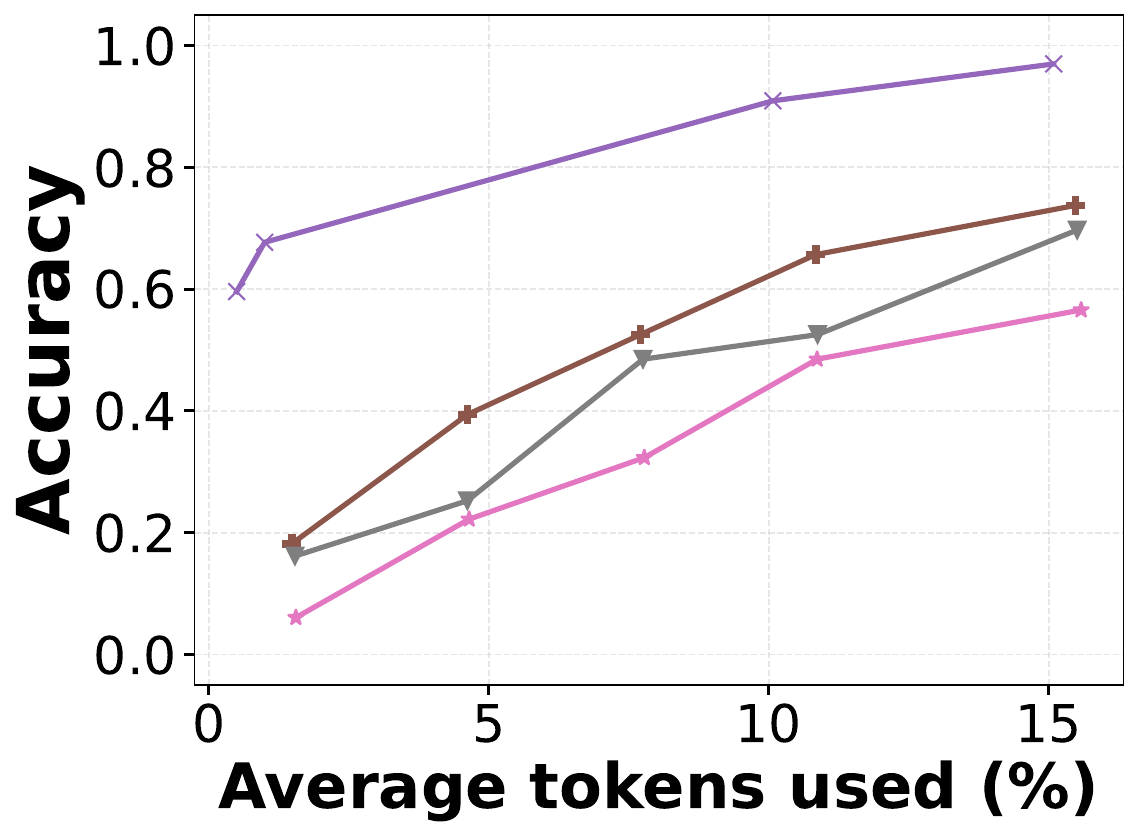}
        \caption{What is the main contribution of the paper}
        \label{fig:lpr-b}
    \end{subfigure}
    \begin{subfigure}[t]{0.24\textwidth}
        \centering
        \includegraphics[width=\linewidth]{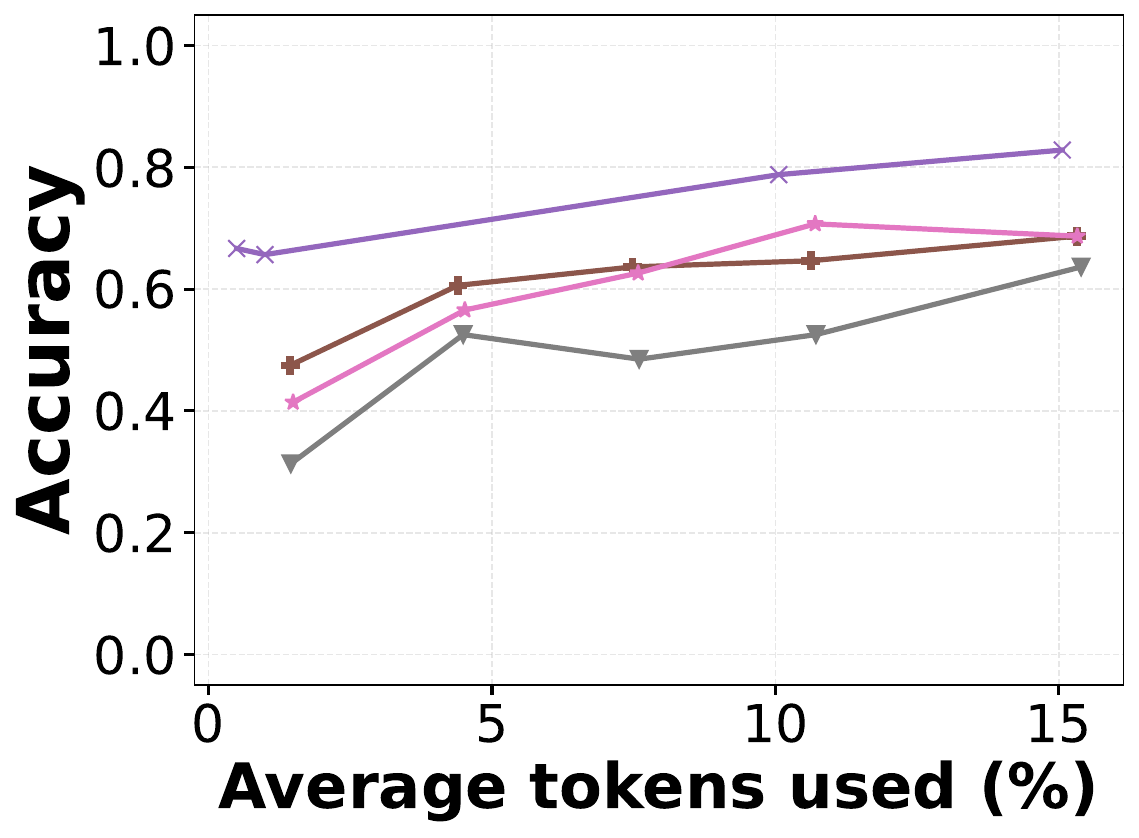}
        \caption{What is the publication year and venue of the paper}
        \label{fig:lpr-c}
    \end{subfigure}
    \begin{subfigure}[t]{0.24\textwidth}
        \centering
        \includegraphics[width=\linewidth]{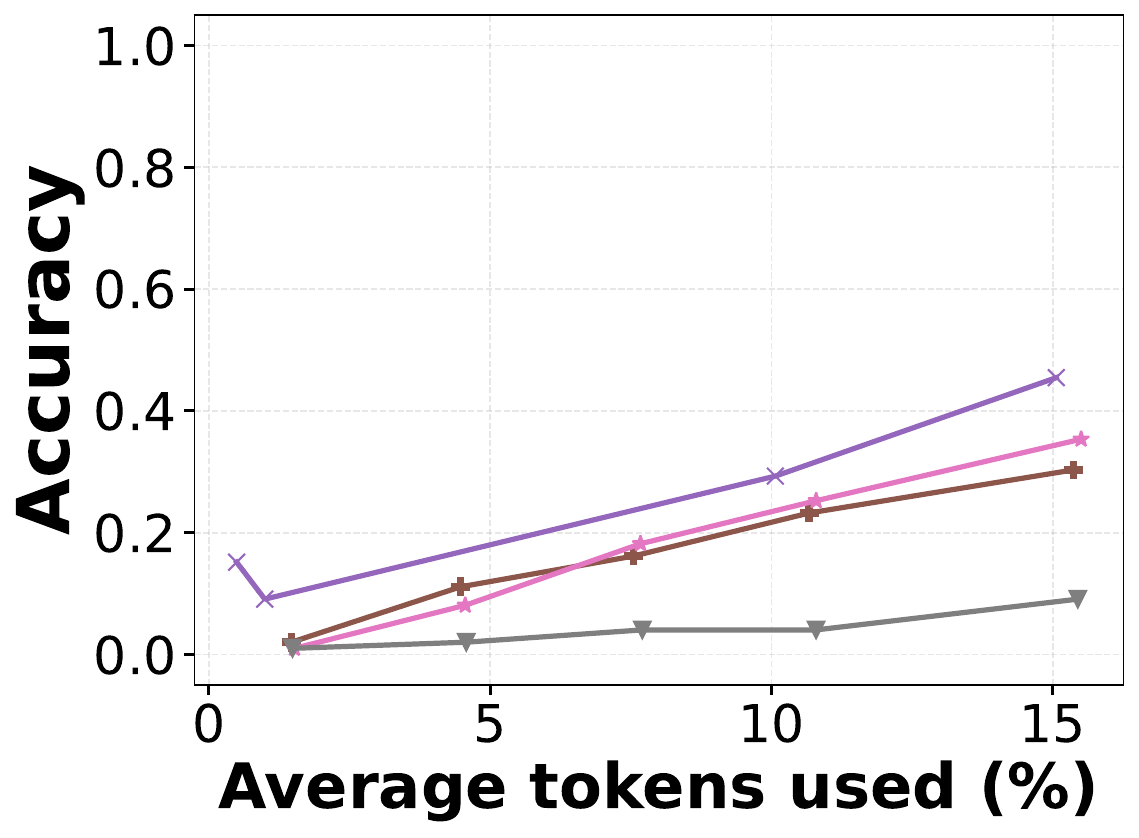}
        \caption{What is the title of the paper}
        \label{fig:lpr-d}
    \end{subfigure}

    \caption{
    \textbf{LLM-judge accuracy on the Paper dataset}.
    Accuracy vs. average token usage over documents for \ourmethod\ and RAG baselines across representative questions.
    }
    \label{fig:line-paper}
\end{figure*}

\begin{figure*}[htbp]
    \centering

    \begin{subfigure}[t]{\linewidth}
        \centering
        \includegraphics[width=\linewidth]{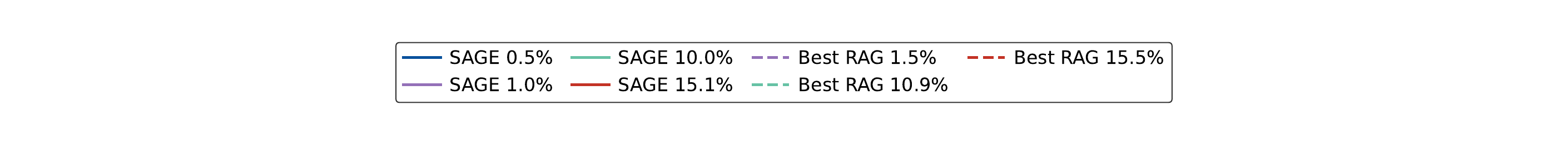}
        \label{fig:cpr-a}
    \end{subfigure}
    \vspace{-20pt}

    \begin{subfigure}[t]{0.23\textwidth}
        \centering
        \includegraphics[width=\linewidth]{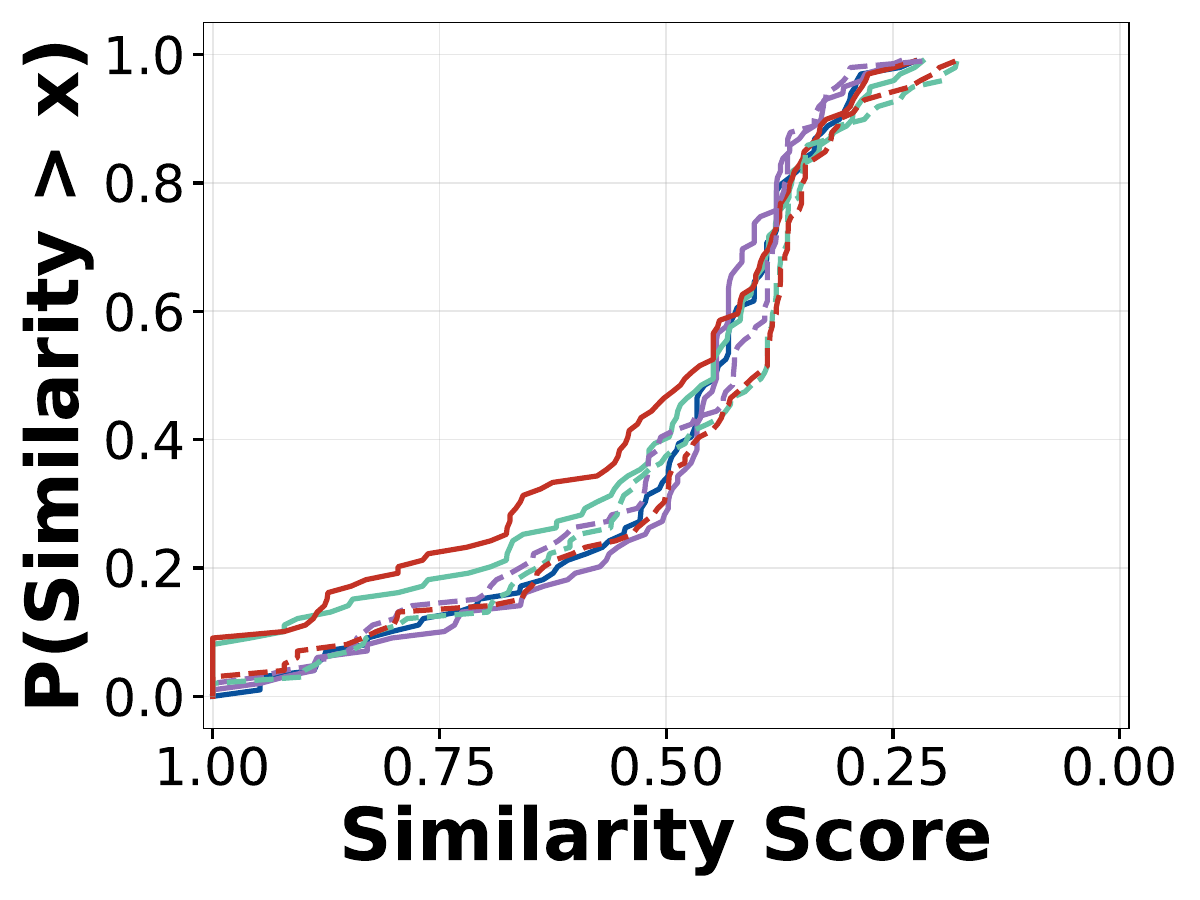}
        \caption{What is the number of authors}
        \label{fig:cpr-c}
    \end{subfigure}
    \begin{subfigure}[t]{0.23\textwidth}
        \centering
        \includegraphics[width=\linewidth]{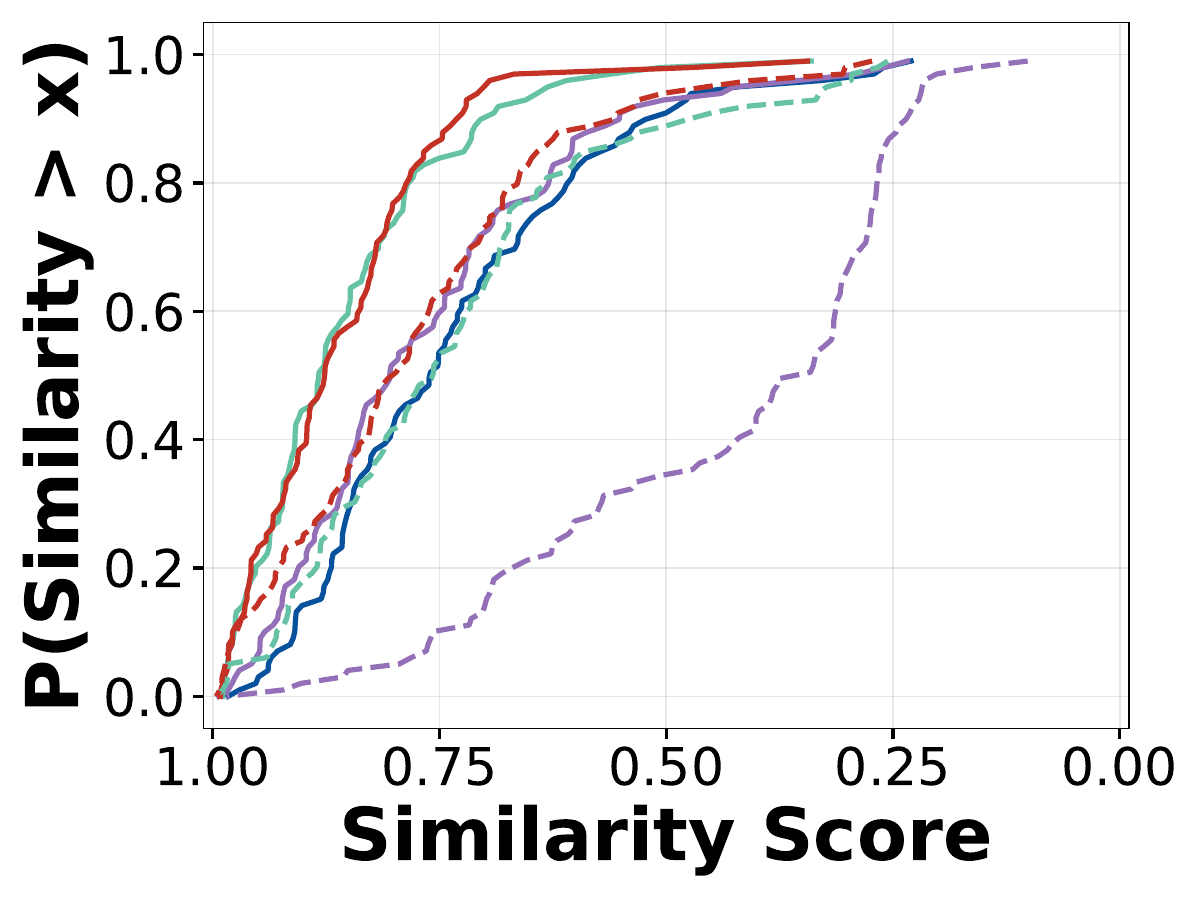}
        \caption{What is the main contribution of the paper}
        \label{fig:cpr-e}
    \end{subfigure}
        \begin{subfigure}[t]{0.23\textwidth}
        \centering
        \includegraphics[width=\linewidth]{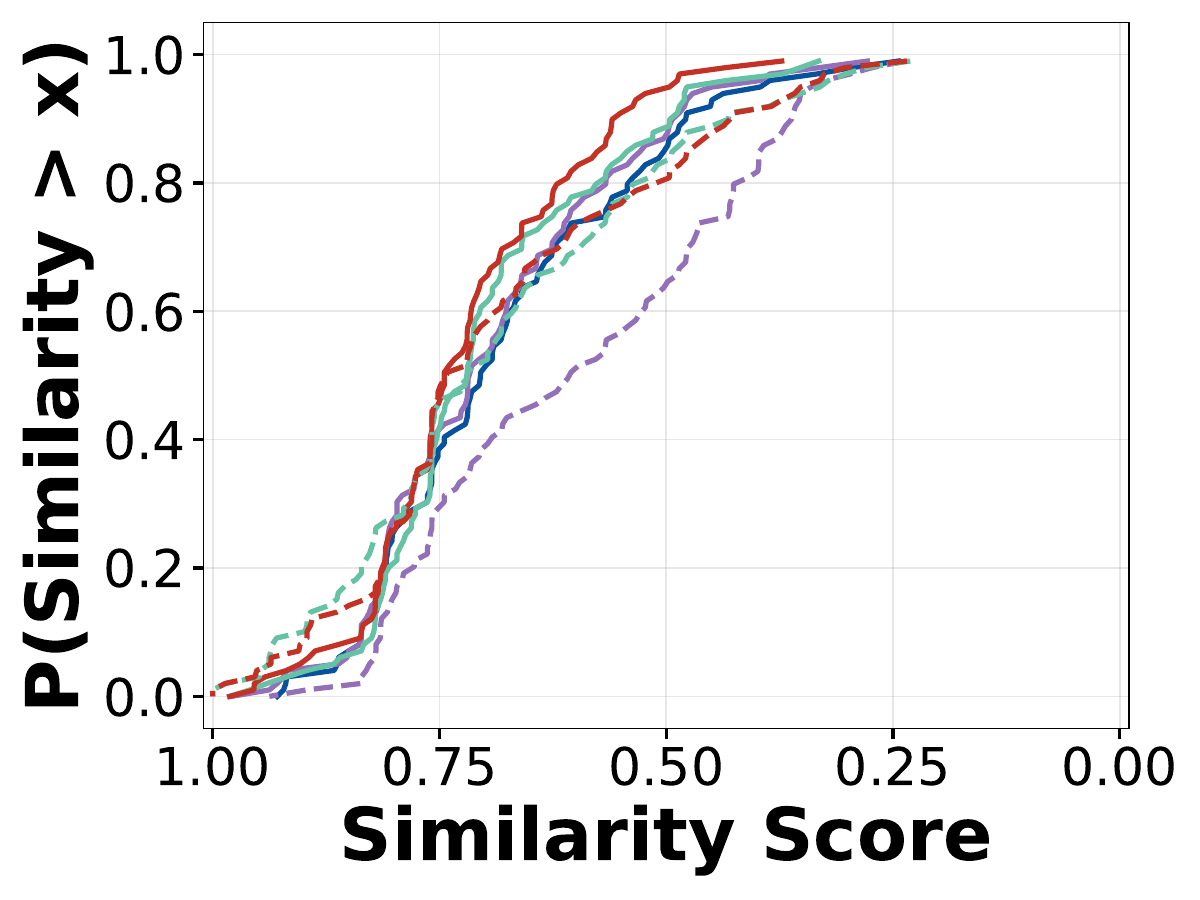}
        \caption{What is the publication year and venue of the paper}
        \label{fig:cpr-b}
    \end{subfigure}
        \begin{subfigure}[t]{0.23\textwidth}
        \centering
        \includegraphics[width=\linewidth]{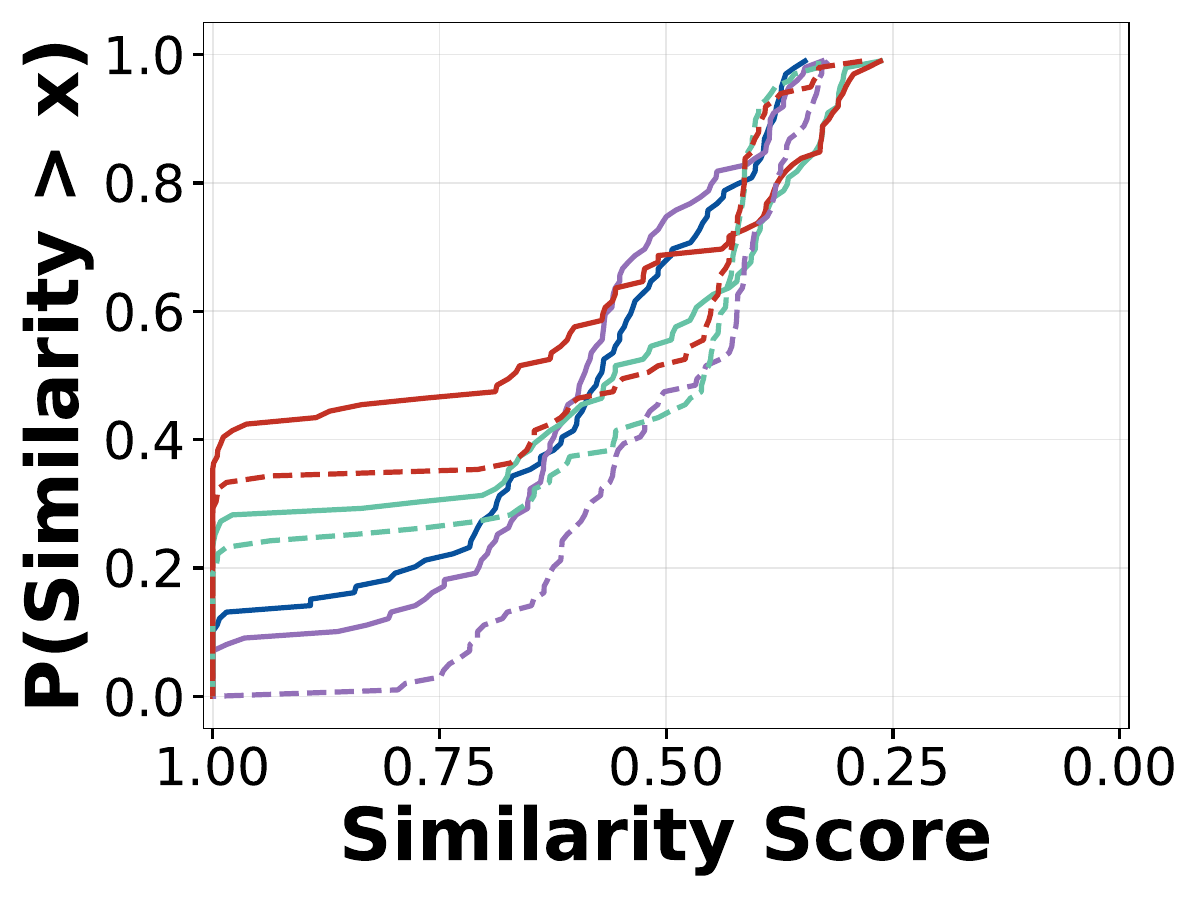}
        \caption{What is the title of the paper}
        \label{fig:cpr-d}
    \end{subfigure}

    \caption{
    \textbf{Semantic similarity CDFs on the Paper dataset.}
CDFs of cosine similarity between generated and ground-truth answers, comparing \ourmethod\ to the best RAG baseline at matched average token usage (including the 0.5\% budget). Results use Qwen3-8B.
    }
    \label{fig:cdf-paper}
\end{figure*}

We next evaluate \ourmethod\ on the \textit{Paper} dataset from ZenDB \citep{lin2024towards}. Compared to QuALITY-hard, Paper pushes context reduction into a more realistic long document setting. Scientific papers are substantially longer and follow a clear section structure. Their questions vary from asking localized metadata, such as venue and year, to open-ended semantic queries like the main contribution of a paper. This combination lets us test whether the advantages of \ourmethod\ extend beyond narrative passages. Concretely, \textit{Paper} answers (Q1), (Q2), (Q3), and (Q7), and serves as evidence for (Q6). Without any dataset-specific tuning, \ourmethod\ remains effective on Paper, indicating that the benefits of attention-guided context reduction extend beyond narrative passages to structurally complex scientific writing.

\subsubsection{LLM-judge accuracy: (Q1) to (Q3)}
\Cref{fig:line-paper} reports LLM-judge accuracy over average token usage for several representative questions. Overall, \ourmethod\ achieves higher accuracy than RAG baselines at similar budgets ((Q1)), and its performance increases smoothly with additional context, indicating that the extra tokens selected by attention are generally helpful rather than distracting.

The difference is most visible on reasoning-intensive questions where evidence is not well captured by keyword similarity, addressing (Q2).  For example, on \emph{``What is the main contribution of the paper?''}, \ourmethod\ improves rapidly with budget and remains consistently above RAG across the displayed range. In other words, as we allow more context, the extra tokens selected by \ourmethod\ more often contain the missing evidence needed to answer the question correctly, whereas retrieving more chunks improves correctness more slowly.

For questions with more localized answers, such as publication year and venue in \Cref{fig:lpr-c}, both methods benefit from additional context, but \ourmethod\ reaches high accuracy with fewer tokens. For metadata queries such as title and author count, \ourmethod\ again achieves consistently stronger accuracy under the same budgets, reflecting more reliable selection of document regions that contain the relevant fields.

We also evaluated larger budgets up to 40\%. However, for most questions, performance flattens out at moderate budgets, with only a small subset continuing to improve at higher budgets. This answers (Q3) and motivates focusing \Cref{fig:line-paper} on the most informative range up to 15\%, which captures the most informative setting where differences between methods are most pronounced.

\begin{figure*}[htbp]
    \centering
    \vspace{-1em}
    \begin{subfigure}[t]{\linewidth}
        \centering
        \includegraphics[width=\linewidth]{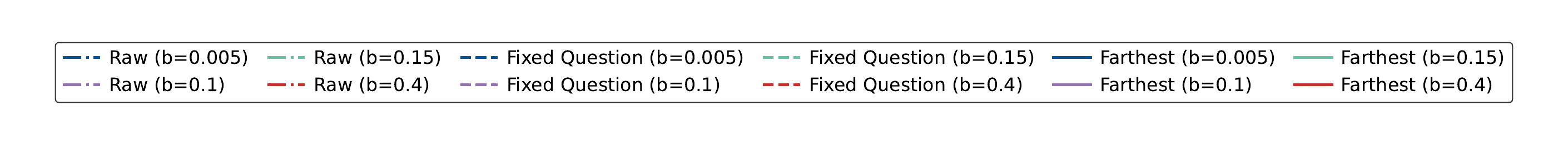}
        \label{fig:p-0}
    \end{subfigure}
    \vspace{-20pt}
    
   \begin{subfigure}[t]{0.25\textwidth}
        \centering
        \includegraphics[width=\linewidth]{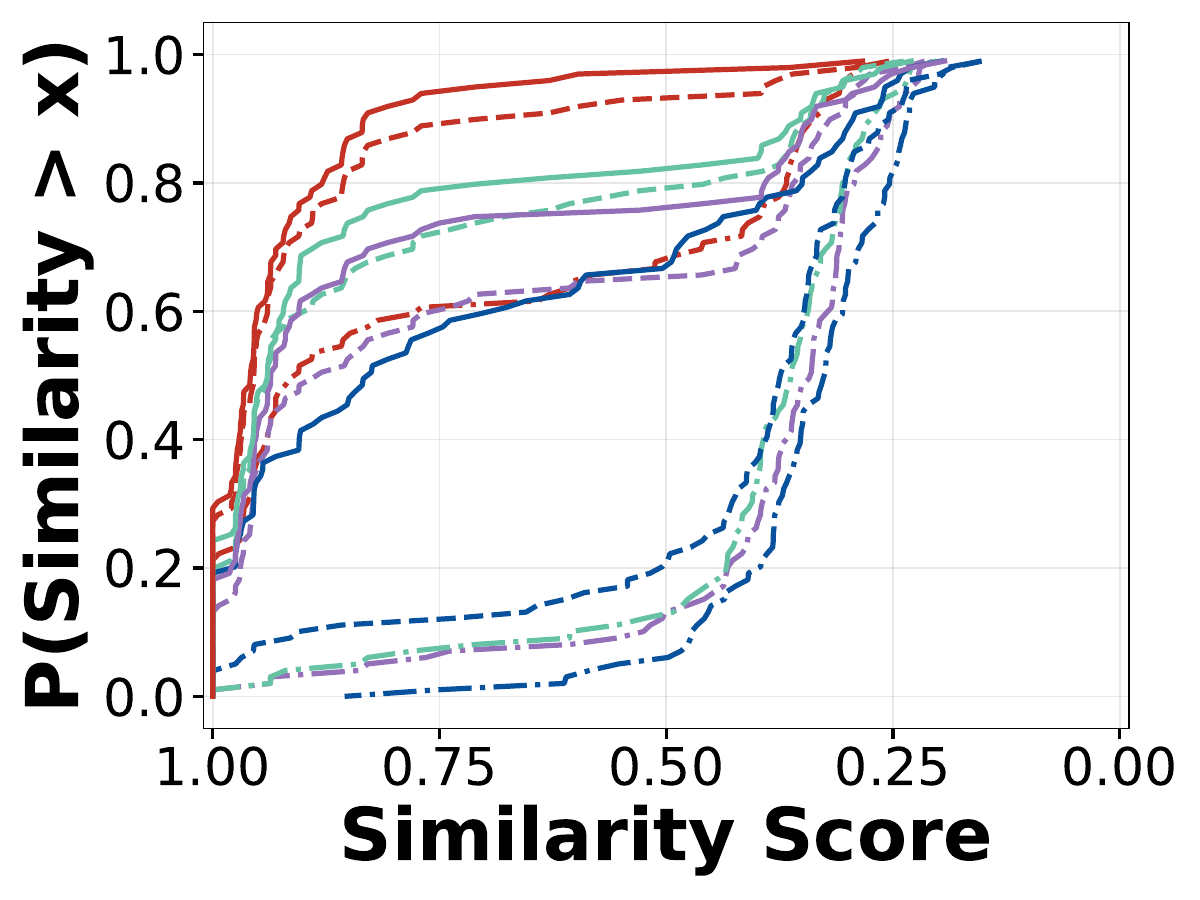}
        \vspace{-1em}
        \caption{what is the authors of the paper}
        \label{fig:p-a}
    \end{subfigure}
    \hspace{50pt}
    \begin{subfigure}[t]{0.25\textwidth}
        \centering
        \includegraphics[width=\linewidth]{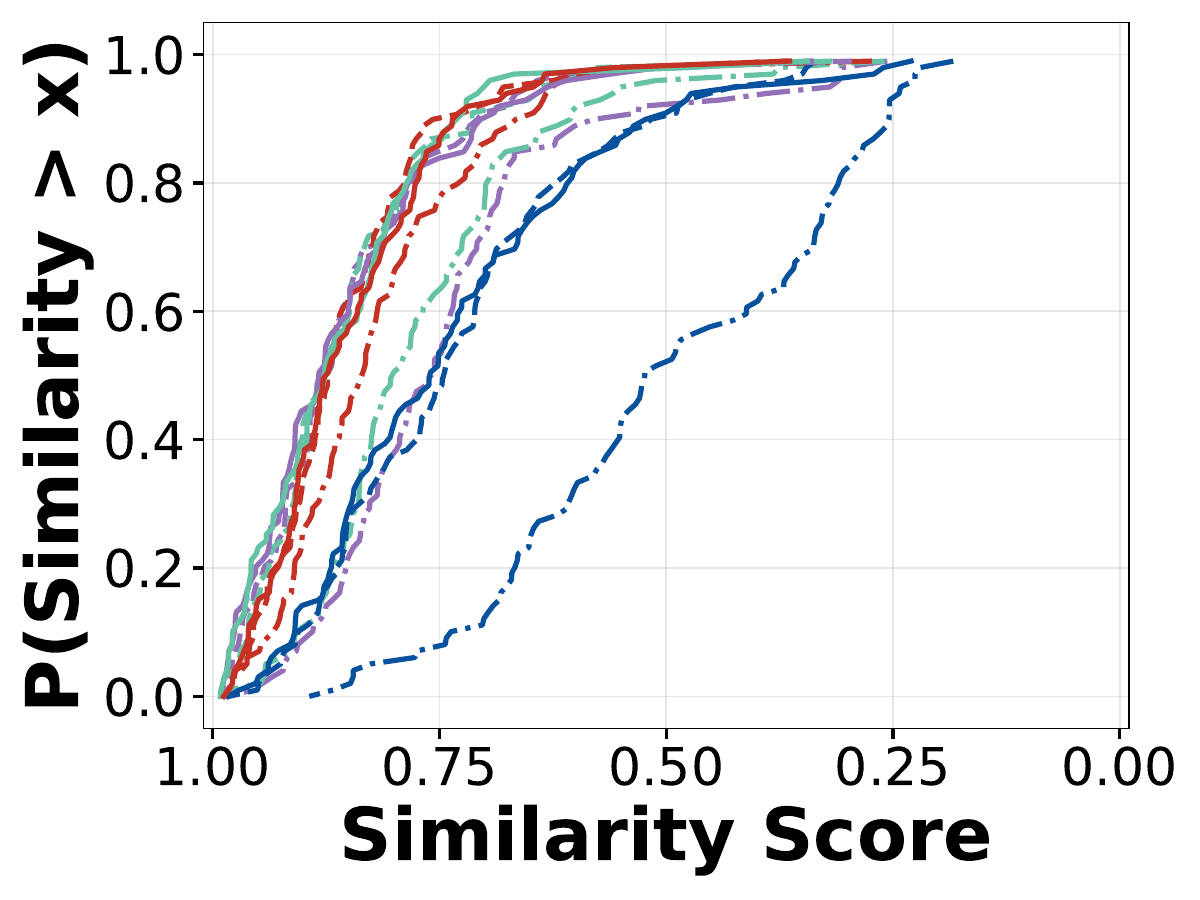}
        \vspace{-1em}
        \caption{what is the main contribution of the paper}
        \label{fig:p-b}
    \end{subfigure}
    \vspace{-1em}
    \caption{
    \textbf{Effect of differential attention on the Paper dataset.}
    Semantic similarity CDFs compare Raw attention with differential attention variants under different token budgets. Experiments are conducted using Qwen3-8B.
    }
    \label{fig:paper-cdf-attention}
\end{figure*}

\subsubsection{Semantic similarity: (Q1) to (Q3)}
We complement judge accuracy with semantic similarity distributions. \Cref{fig:cdf-paper} plots CDFs of cosine similarity between generated answers and ground truth references. For a similarity threshold $x$, the y-axis reports the fraction of questions whose similarity exceeds $x$. Curves that lie higher indicate better overall answer quality. We plot the similarity threshold from 1 to 0 on the x-axis to make high similarity behavior easier to compare. We also evaluated ROUGE. It shows the same overall trends. For clarity, we report cosine similarity.

Across representative questions, \ourmethod\ dominates the best performing RAG baseline in the middle-to-high similarity region, reinforcing (Q1) by showing improvements in overall answer quality rather than only binary pass rates. The largest gains again appear on open-ended reasoning questions such as ``main contribution''. This supports our answer to (Q2). Under the same budget, \ourmethod\ produces a higher fraction of answers that closely match the reference semantics, while increasing retrieval depth yields smaller shifts in the distribution. The CDFs also answer (Q3): increasing the budget generally shifts curves upward, but the improvements flatten out at larger budgets. 

Taken together, the judge and similarity views agree that \ourmethod\ converts budget into answer quality more effectively than retrieval, especially for reasoning-intensive semantic questions.

\subsubsection{Different Differential Attention Variant (Q7)}
We study the differential attention variant on \textit{Paper}. \Cref{fig:paper-cdf-attention} compares raw attention with fixed contrast query and farthest query. Across budgets and representative questions, both differential variants shift the similarity CDFs upward relative to raw attention, meaning they produce a larger fraction of high-quality answers under the same token budget. This directly answers (Q7): differential attention strengthens context selection by filtering out noise and producing cleaner, query-specific relevance scores.

Between the two variants, farthest question performs best overall. Its CDF curves dominate the fixed contrast variant across most similarity thresholds, with especially clear gains on reasoning questions (\Cref{fig:p-b}). In practice, prefer an informative contrast question when available. Otherwise, use fixed contrast, which still delivers strong performance.

\subsection{Notice}
\label{sec:results-notice}
We next evaluate on the \textit{Notice} dataset from ZenDB \citep{lin2024towards}. Notice documents have less variation in length, which enables more controlled comparisons with aligned token usage between \ourmethod\ and RAG. Extremely small budgets that fall below a single retrieval chunk are studied on this dataset. As Notice has a markedly different document style, the results below provide additional evidence for (Q6), showing \ourmethod remains effective without specific tuning.

\begin{figure*}[htbp]
    \centering

    \begin{subfigure}[t]{\linewidth}
        \centering
        \includegraphics[width=\linewidth]{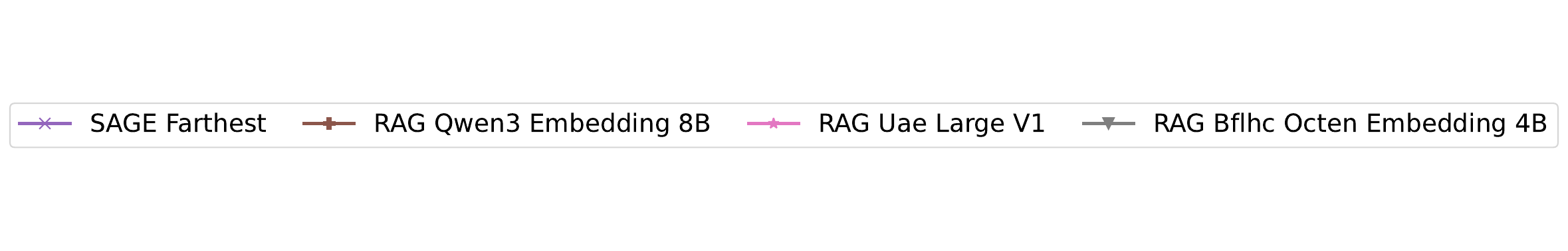}
        \label{fig:lnr-0}
    \end{subfigure}
    \vspace{-40pt}

    \begin{subfigure}[t]{0.24\textwidth}
        \centering
        \includegraphics[width=\linewidth]{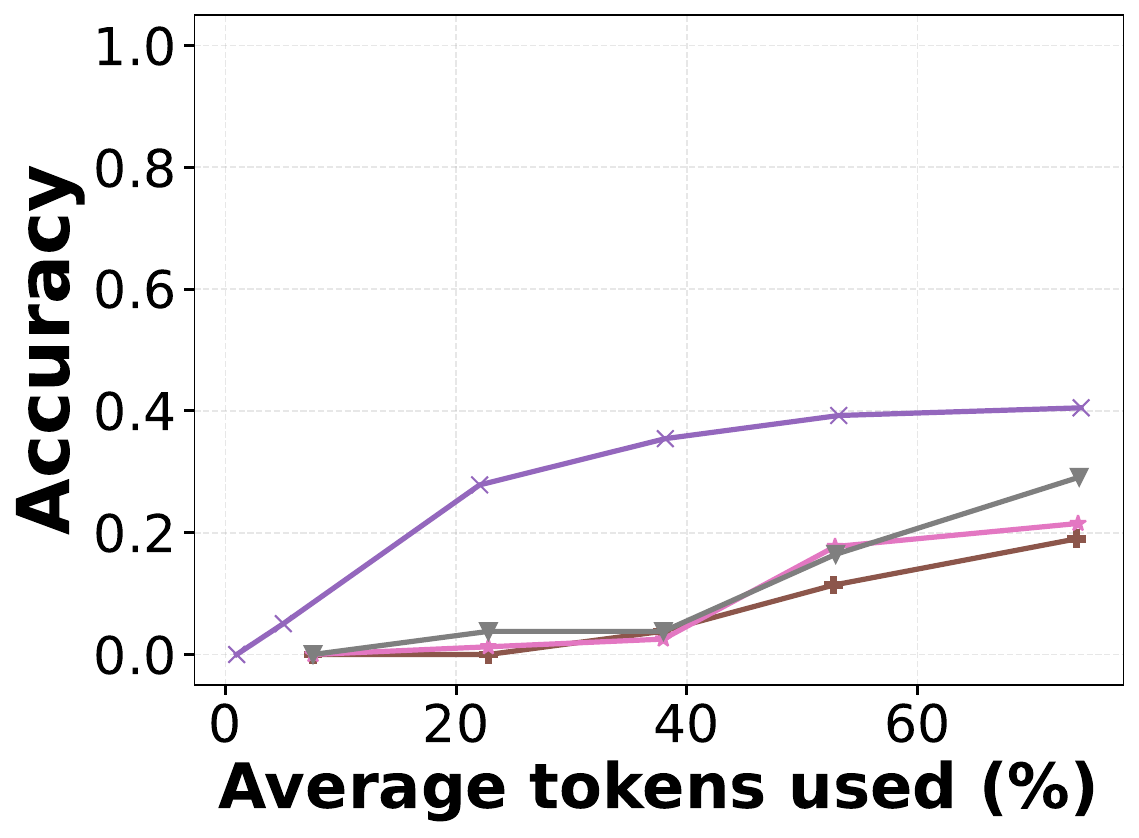}
        \caption{What are the violation items}
        \label{fig:lnr-a}
    \end{subfigure}
    \begin{subfigure}[t]{0.24\textwidth}
        \centering
        \includegraphics[width=\linewidth]{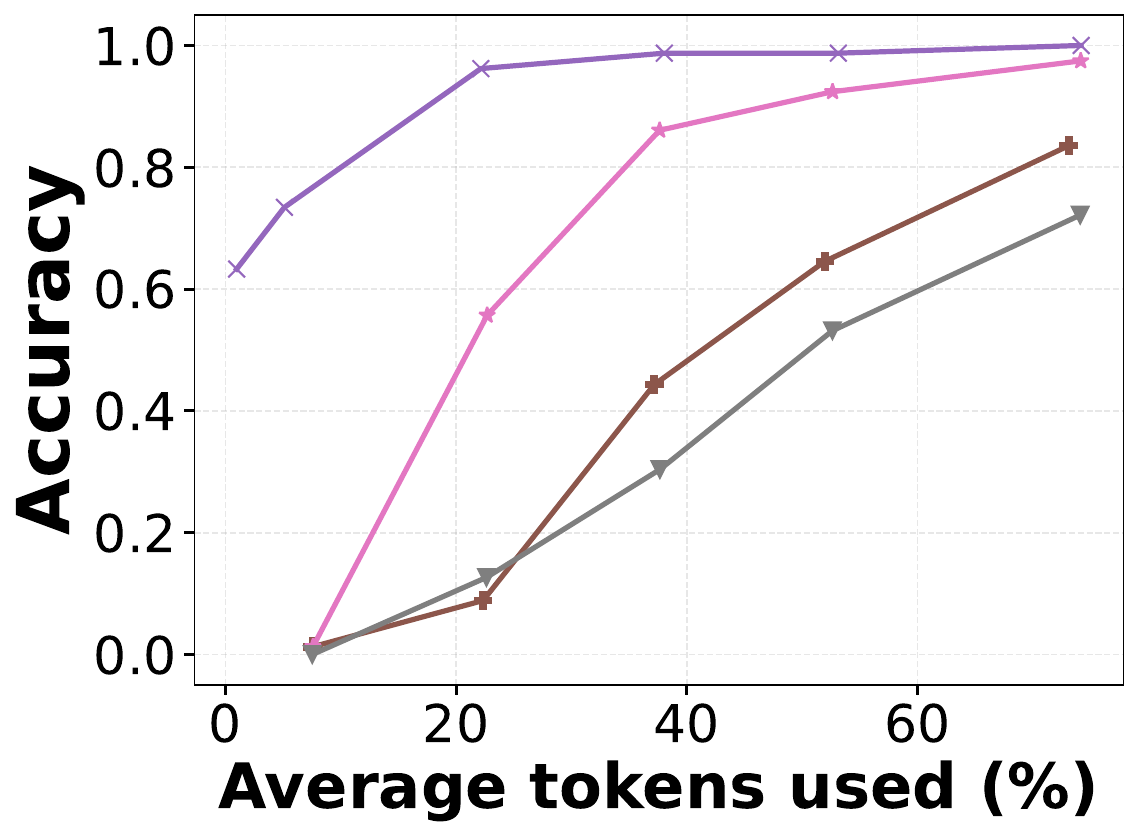}
        \caption{What is the date of the notice}
        \label{fig:lnr-b}
    \end{subfigure}
    \begin{subfigure}[t]{0.24\textwidth}
        \centering
        \includegraphics[width=\linewidth]{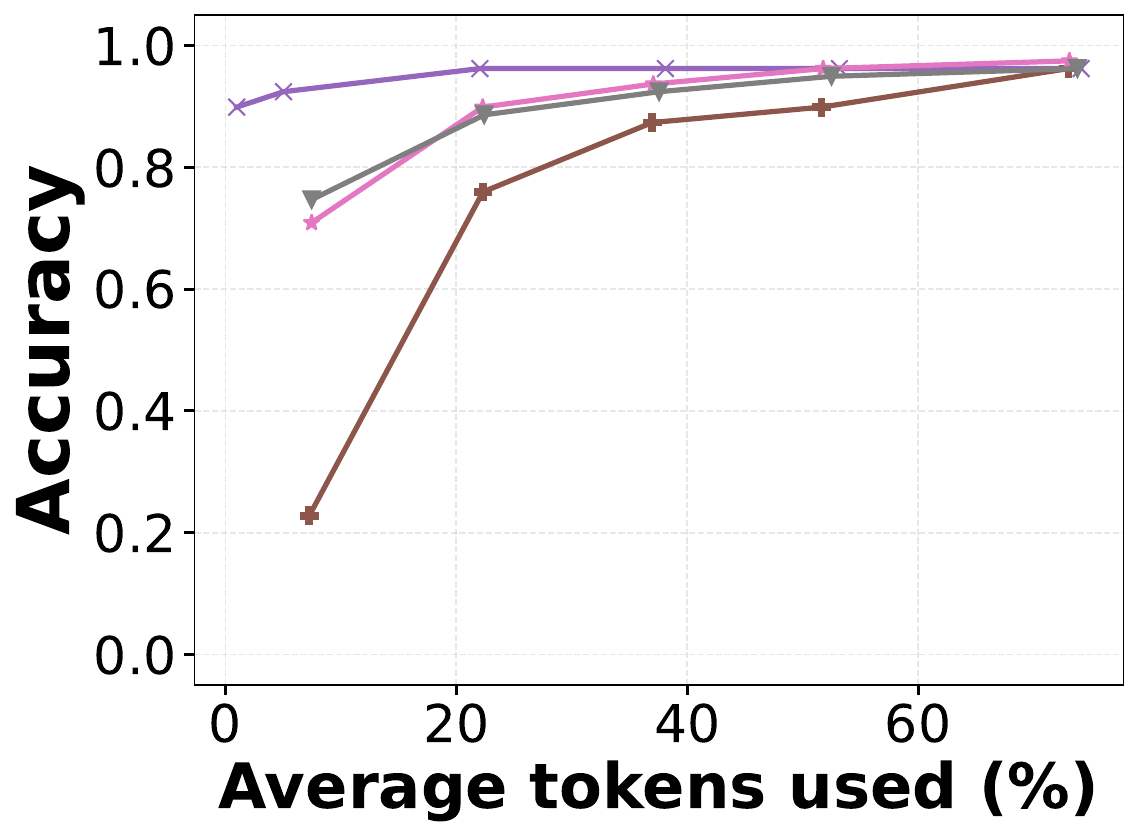}
        \caption{What is the name of the company}
        \label{fig:lnr-c}
    \end{subfigure}
    \begin{subfigure}[t]{0.24\textwidth}
        \centering
        \includegraphics[width=\linewidth]{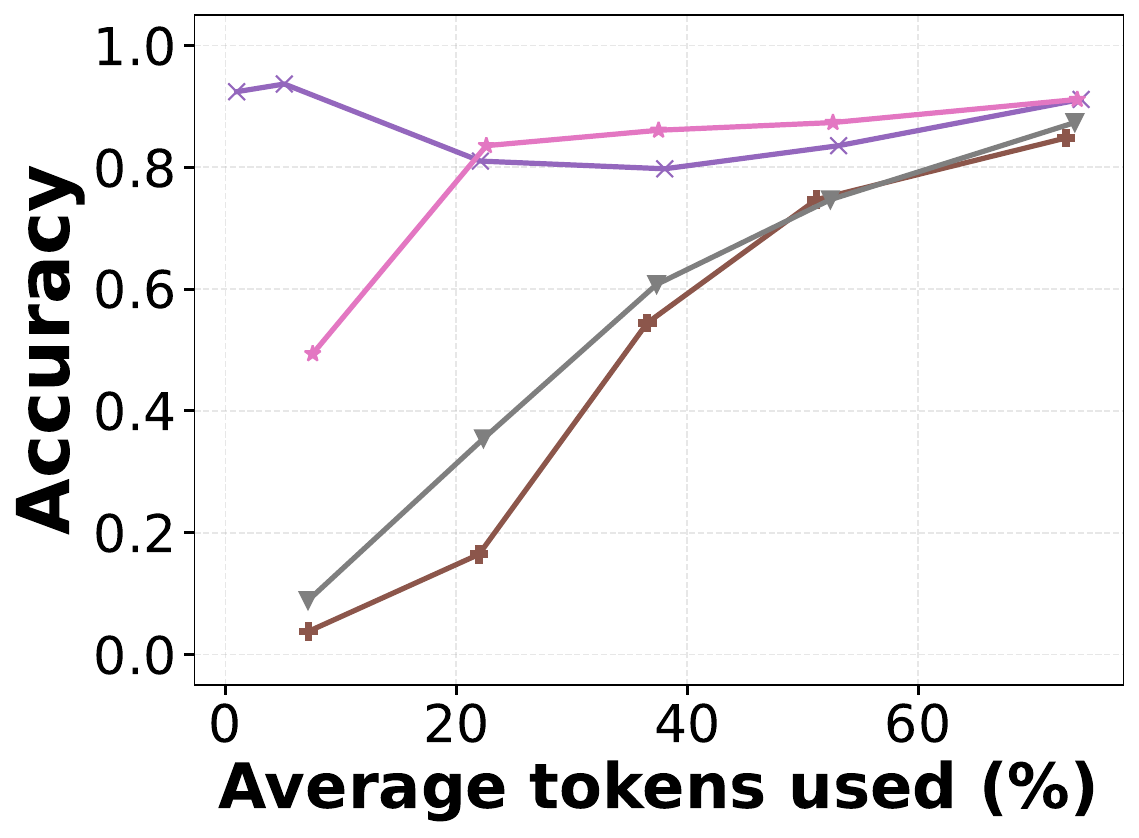}
        \caption{What are the state abbreviation and ZIP code of the company}
        \label{fig:lnr-d}
    \end{subfigure}
    
    \caption{
    \textbf{LLM-judge accuracy on the Notice dataset.}
    Accuracy versus average token usage for \ourmethod\ and RAG baselines across representative questions. \ourmethod\ supports budgets smaller than a single RAG chunk. For question (d), accuracy decreases as token usage increases; we discuss this failure mode in \Cref{sec:results-notice}.
    }
    \label{fig:line-notice}
\end{figure*}

\subsubsection{Accuracy and semantic quality under aligned budgets}
\Cref{fig:line-notice} reports LLM-judge accuracy versus average token usage for representative questions, and \Cref{fig:cdf-notice} complements this view with semantic similarity distributions. Across questions, \ourmethod\ matches or outperforms strong RAG baselines at aligned token usage, while also supporting substantially smaller budgets (Q1).

\subsubsection{Budgets below a single retrieval chunk (Q8)}
Notice highlights a practical limitation of chunk-based retrieval. Even retrieving a single chunk incurs a fixed token cost, which limits its flexibility under tight budgets. In contrast, \ourmethod\ selects evidence at finer granularity and remains effective even when the budget is smaller than a single retrieved chunk. This answers (Q8): \ourmethod\ can operate at extremely low budgets that chunk-based RAG cannot meaningfully support.

\begin{figure*}[htbp]
    \centering
    \begin{subfigure}[t]{\linewidth}
        \centering
        \includegraphics[width=\linewidth]{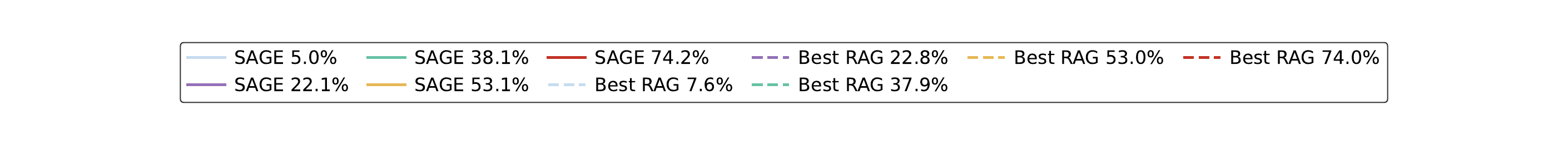}
        \label{fig:cnr-0}
    \end{subfigure}
    \vspace{-20pt}

    \begin{subfigure}[t]{0.24\textwidth}
        \centering
        \includegraphics[width=\linewidth]{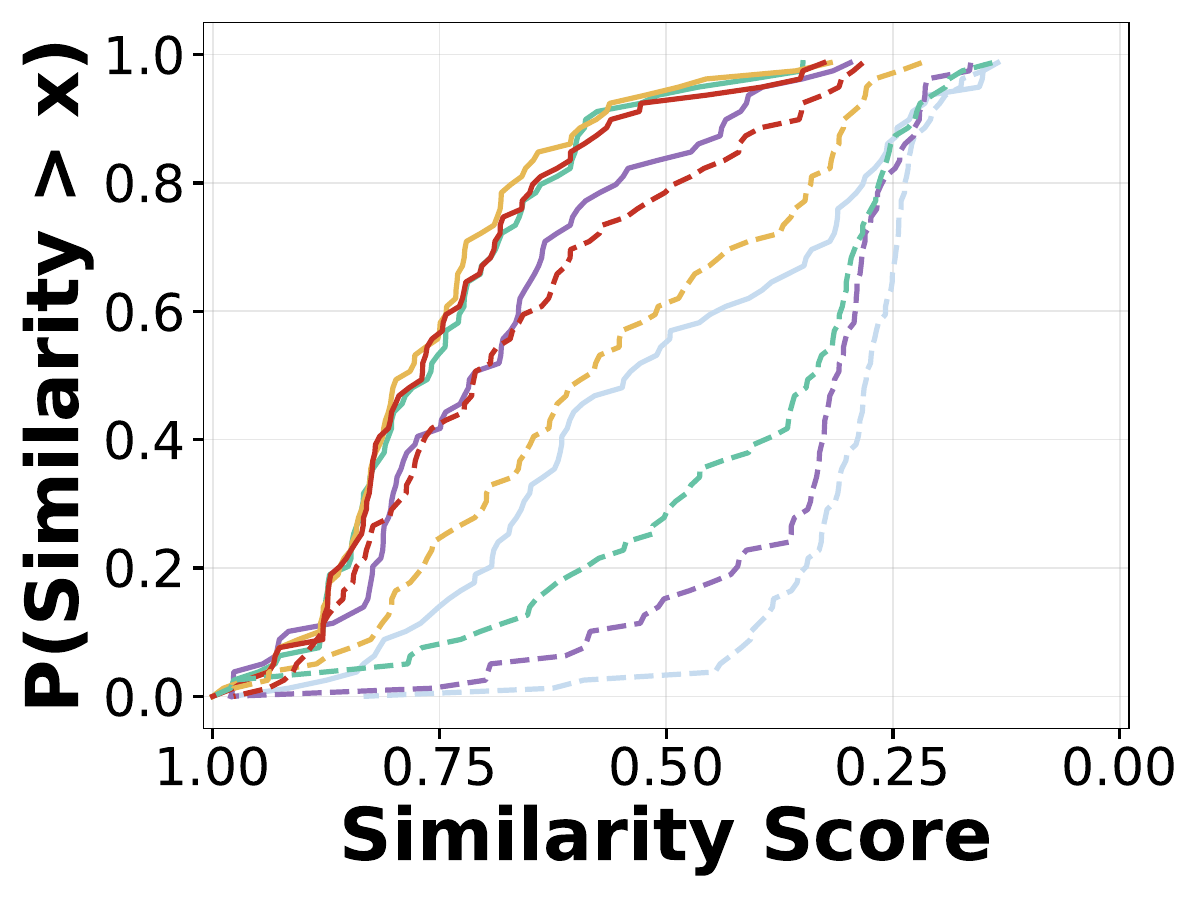}
        \caption{What are the violation items}
        \label{fig:cnr-a}
    \end{subfigure}
    \begin{subfigure}[t]{0.24\textwidth}
        \centering
        \includegraphics[width=\linewidth]{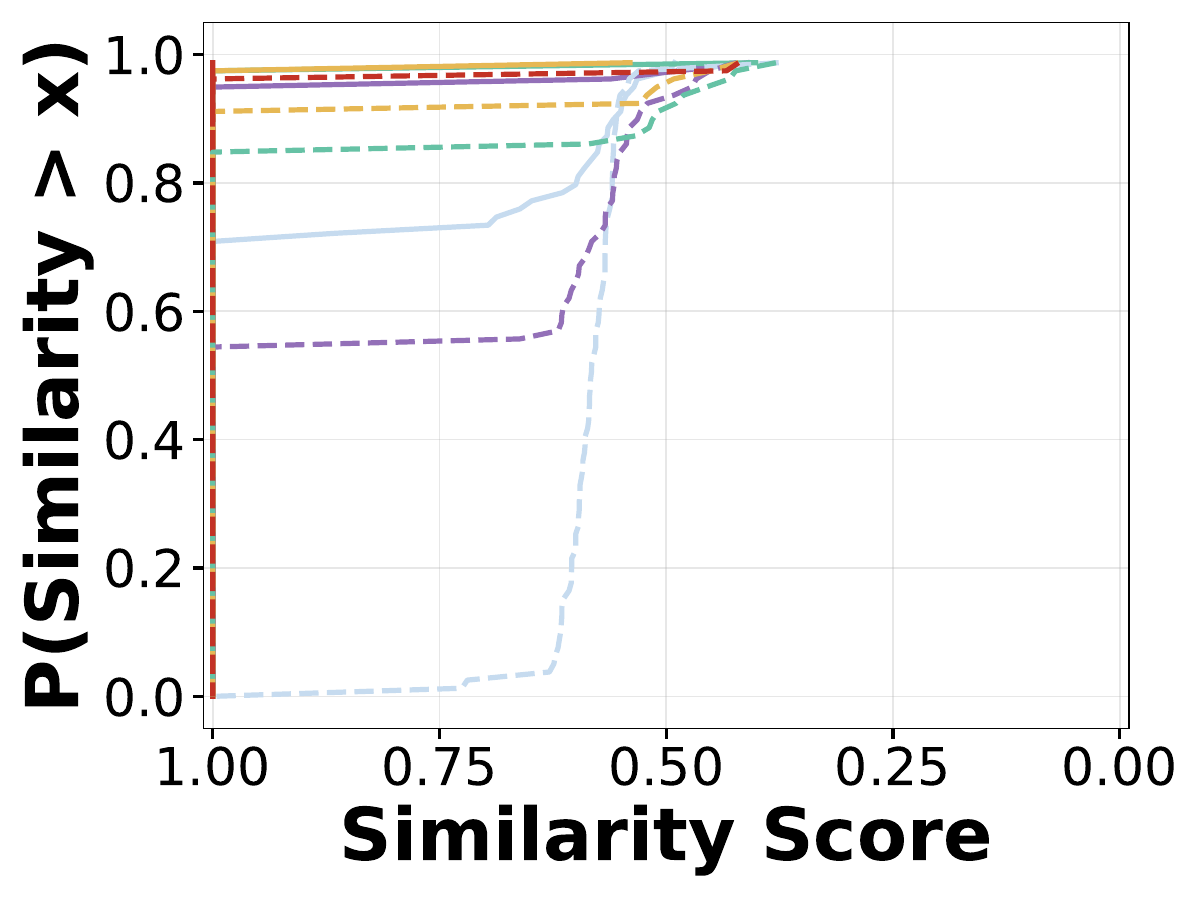}
        \caption{What is the date of the notice}
        \label{fig:cnr-b}
    \end{subfigure}
    \begin{subfigure}[t]{0.24\textwidth}
        \centering
        \includegraphics[width=\linewidth]{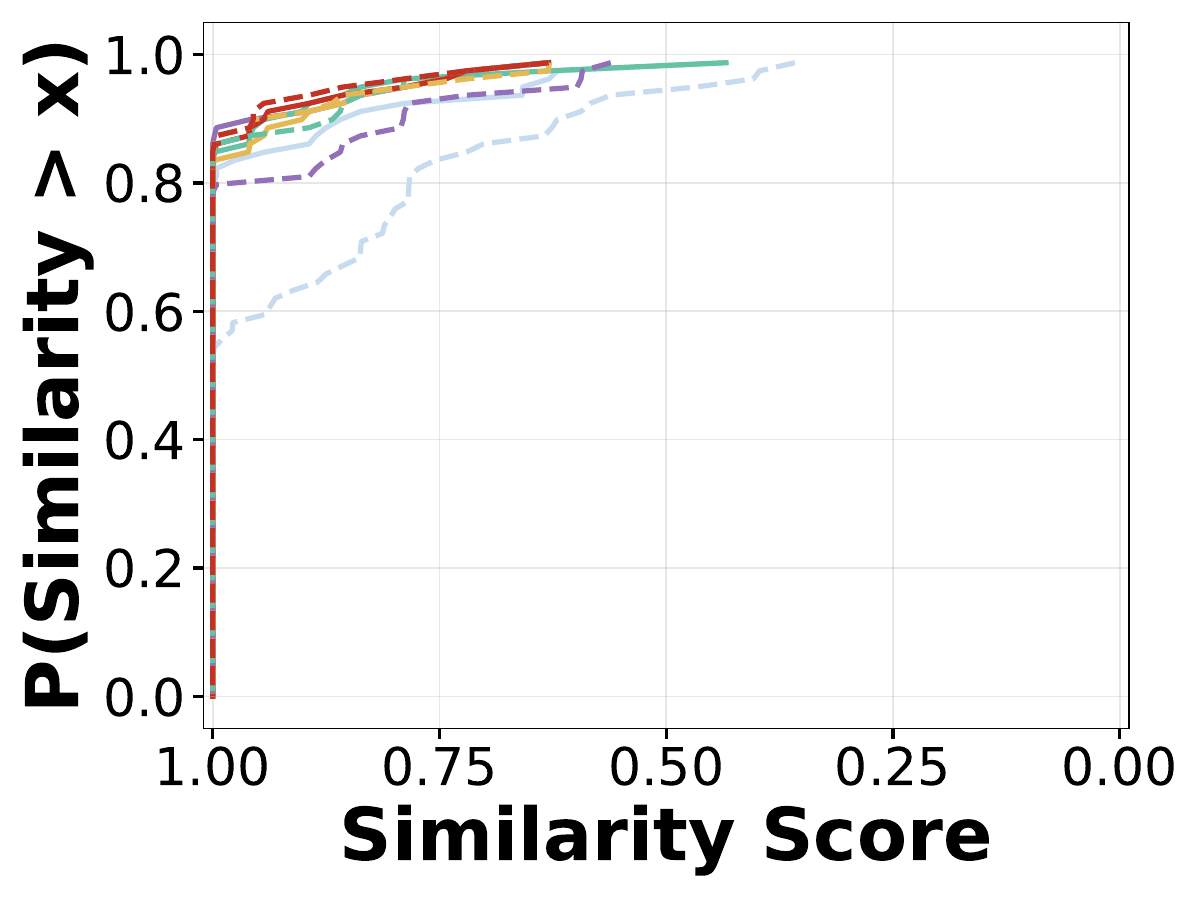}
        \caption{What is the name of the company}
        \label{fig:cnr-c}
    \end{subfigure}
    \begin{subfigure}[t]{0.24\textwidth}
        \centering
        \includegraphics[width=\linewidth]{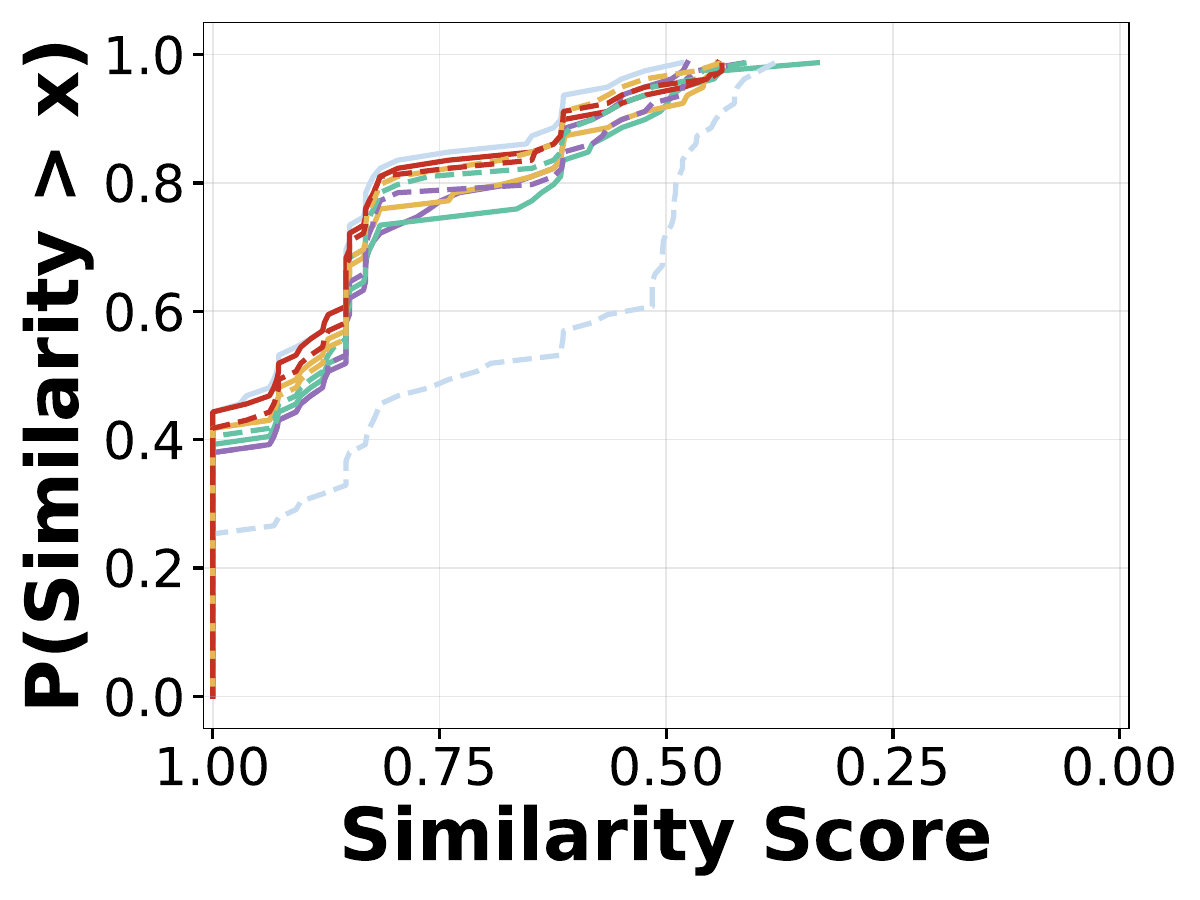}
        \caption{What are the state abbreviation and zip code of the company}
        \label{fig:cnr-d}
    \end{subfigure}
    
    \caption{
    \textbf{Semantic similarity CDFs on the Notice dataset.}
    CDFs compare \ourmethod\ and the best performing RAG baseline under matched average token budgets across representative questions. Results are obtained using Qwen3-8B for attention estimation.
    }
    \label{fig:cdf-notice}
\end{figure*}

\paragraph{A failure mode at larger budgets.}
An instructive exception appears for \emph{``What are the state abbreviation and ZIP code of the company?''} in \Cref{fig:lnr-d}. Here, \ourmethod\ performs best at small budgets, but accuracy decreases as the budget grows and approaches the RAG baselines. The same trend appears in the similarity CDF (\Cref{fig:cnr-d}). This behavior reflects a limitation of greedy attention-based selection in repetitive documents. With a small budget, \ourmethod\ typically captures the query-relevant ZIP code. As the budget increases, additional sections containing other ZIP codes may be included, introducing competing candidates and reducing precision. More broadly, this highlights that adding more context is not always beneficial when documents contain repeated entities and near duplicate fields.

\subsection{AIT-QA}
\label{sec:results-aitqa}

\begin{table}[t]
\centering
\caption{AIT-QA results comparing full-table RAG and Top-4 row selection methods.}
\label{tab:aitqa-top4}
\begin{tabular}{lcc}
\toprule
\textbf{Method} & \textbf{Value} & \textbf{Avg. Row Usage} \\
\midrule
Full Table             & 0.88 & 100\% \\
\hline
\ourmethod           & 0.87 & 51\%  \\
\bottomrule
\end{tabular}
\end{table}

We finally evaluate whether attention-guided selection extends beyond free-form text to semi-structured inputs, addressing (Q9), using the \textit{AIT-QA} dataset \citep{katsis2021aitqa}. Each document in AIT-QA is a structured table represented in JSON format. As a result, neither standard RAG nor token-based context reduction can be applied directly. If a table is split into text chunks or partial windows would break row–column structure and substantially hinder reasoning.

To handle this setting, we adapt \ourmethod\ to operate at the \emph{row level}. We compute query-aware attention over the full table in its original JSON representation. For each row, we aggregate token-level attention into a single relevance score and then select the top-$k$ rows. The reduced prompt includes only the selected rows along with the column headers.

\Cref{tab:aitqa-top4} compares this row-selection variant with a full-table baseline. Selecting only the top rows achieves accuracy comparable to using the full table, while substantially reducing the amount of table content passed to the generation model. This answers (Q9), \ourmethod\ can preserve task accuracy on semi-structured tables while enabling budget-aware context reduction.

These findings show that \ourmethod\ remains effective beyond unstructured text, highlighting its potential as a flexible, budget-aware context selection framework that transfers across data types without additional fine-tuning (Q6).

\section{Conclusion}
\label{sec:conclude}


We presented \ourmethod, a training-free, plug-and-play context reduction framework that leverages a lightweight local LLM to extract highly compact, fine-grained, and query-relevant subsets from long documents under strict token budgets. Across four complementary benchmarks, \ourmethod\ consistently outpaces strong RAG baselines in token efficiency, demonstrating its largest gains on complex, reasoning-intensive questions. Furthermore, by utilizing differential attention, the framework remains highly effective even on substantially longer, noisy scientific papers. Beyond standard text, our results on AIT-QA confirm that \ourmethod\ naturally extends to structured inputs, preserving critical tabular data while drastically reducing the exposed context. Moving forward, we aim to enhance disambiguation mechanisms for complex structured data, integrate multi-stage retrieval, and expand \ourmethod\ to process richer multimodal documents.



\bibliographystyle{ACM-Reference-Format}
\bibliography{ref}

\end{document}